\documentclass[]{aastex631}
\usepackage{amsmath}

\AtBeginDocument{%

}

\usepackage{xcolor}

\begin{document}

% \title{The Inversion of Initial Water Content and Age Estimation for GJ 486b}
% \title{Estimating Stellar Age Based on Planetary Water Content Inversion: The Case of GJ 486b}
% \title{\textcolor{red}{Water Content Inversion and Age Estimation with Planetary Formation Dataset for GJ 486b}}
\title{
Inversion of Hydrogen-rich Atmosphere and Water Content for GJ 486b
}

\correspondingauthor{Jianheng Guo}
\email{guojh@ynao.ac.cn}

\author[0009-0004-7473-1727]{Junda Zhou}
\affiliation{Yunnan Observatories, Chinese Academy of Sciences, Kunming 650216, China}
\affiliation{University of Chinese Academy of Sciences, Beijing 100049, China}
\affiliation{International Centre of Supernovae, Yunnan Key Laboratory, Kunming 650216, China}

\author[0009-0009-8188-5632]{Zhenyang Huang}
\affiliation{Xinjiang Astronomical Observatory, Chinese Academy of Sciences, Urumqi 830011, China}
\affiliation{University of Chinese Academy of Sciences, Beijing 100049, China}

\author[0000-0003-0707-3213]{Di-Chang Chen}
\affiliation{School of Physics and Astronomy, Sun Yat-sen University, Zhuhai, China}
\affiliation{Center of CSST in the Great Bay Area, Sun Yat-sen University, Zhuhai, China}

\author[0000-0002-8869-6510]{Jianheng Guo}
\affiliation{ International Centre of Supernovae (ICESUN), Yunnan Key Laboratory of Supernova Research, Yunnan Observatories, Chinese Academy of Sciences (CAS), Kunming 650216, China}
\affiliation{University of Chinese Academy of Sciences, Beijing 100049, China}

% \author{Junda Zhou}

% \author[0009-0009-8188-5632]{Zhenyang Huang}

% \author{Jianheng Guo}

% \author[0000-0002-0786-7307]{Greg J. Schwarz}
% \affiliation{American Astronomical Society \\
% 1667 K Street NW, Suite 800 \\
% Washington, DC 20006, USA}

% \author{August Muench}
% \affiliation{American Astronomical Society \\
% 1667 K Street NW, Suite 800 \\
% Washington, DC 20006, USA}

% \collaboration{20}{(AAS Journals Data Editors)}

% \author{F.X Timmes}
% \affiliation{Arizona State University}
% \affiliation{AAS Journals Associate Editor-in-Chief}

% \author{Amy Hendrickson}
% \altaffiliation{AASTeX v6+ programmer}
% \affiliation{TeXnology Inc.}

% \author{Julie Steffen}
% \affiliation{AAS Director of Publishing}
% \affiliation{American Astronomical Society \\
% 1667 K Street NW, Suite 800 \\
% Washington, DC 20006, USA}



\begin{abstract}

GJ~486b is a close-in planet orbiting an M dwarf and is therefore expected to have undergone strong atmospheric escape.
Motivated by theoretical and observational studies on the constraints of its water and atmosphere, we investigate which combinations of an primordial hydrogen-rich atmosphere and water inventory could fit the current water content implied by bulk density measurements.
We model the atmosphere escape using VPLanet, following the loss of an initial hydrogen-rich atmosphere and the subsequent escape of a water-dominated atmosphere.
By scanning a broad parameter space across different stellar ages, we invert for the initial hydrogen-rich atmospheric mass and water inventory consistent with the current constraints.
Our results reveal a strong degeneracy between the water reservoir and the initial hydrogen-rich atmosphere.
Even a modest hydrogen-rich atmosphere can significantly delay early escape of the water and reduce the water inventory required to reproduce the current water content.
We also find that the inferred initial conditions are also strongly age dependent.
Incorporating a planet formation dataset as a prior, we derive a probabilistic constraint on the host star age, yielding an expected age of $2.90^{+2.47}_{-2.27}$~Gyr, which is consistent with the results obtained from other methods to determine M dwarf ages.

\end{abstract}

% \keywords{Classical Novae (251) --- Ultraviolet astronomy(1736) --- History of astronomy(1868) --- Interdisciplinary astronomy(804)}

\section{Introduction} \label{sec:intro}

Since the discovery of the first exoplanet in 1992, more than 5,900 exoplanets have been discovered\footnote{https://exoplanetarchive.ipac.caltech.edu/}.
For a long time, Jupiter-like planets were frequently discovered because of their large size and thick atmosphere.
The first terrestrial planet, CoRoT-7b, was discovered by CoRoT space mission \citep{2009A&A...506..287L}.
Since then, a large number of terrestrial planets have been discovered and some of them are considered potentially habitable planets\footnote{https://phl.upr.edu/hwc}.
Among the main sequence stars in the solar neighborhood, M dwarfs are the most abundant \citep{2013AJ....146...99C}.
Planets orbiting M dwarfs are more easily detected by transit photometry due to their small radii \citep{2007AsBio...7...30T}.
% Furthermore, JWST has made detailed observations of the atmospheres of terrestrial planets possible \citep{lustig2019detectability, 2021PASP..133e4401G}.
% Furthermore, since its launch, JWST has enabled the sensitive atmospheric investigations of nearby terrestrial exoplanets, placing meaningful constraints on the presence and possible composition of atmospheres on several rocky planets \citep{2023NatAs...7.1317L,2023Natur.618...39G,2023Natur.620..746Z}.
Furthermore, the launch and application of JWST have made it possible to investigate the nearby terrestrial exoplanets atmospheres and provide significant constraints on the composition of atmospheres on several rocky planets \citep{2023NatAs...7.1317L,2023Natur.618...39G,2023Natur.620..746Z}.
Thus, studying these planetary systems can provide the signals of atmosphere, including the chemical compositions and dynamical characters.
This can enhance our understanding of Earth's unique characteristics as the only known rocky planet that harbors life.

Terrestrial planets orbiting M star can accrete a hydrogen-rich envelope during their formation phase \citep{2016ApJ...825...86S, 2022NatAs...6.1296K}, which is recognized as the primary atmosphere.
% The hydrogen-rich atmosphere can get rid of the bind of the planet to escape to space owing to the thermal energy of planet, intense stellar XUV radiation and stellar tidal forces \citep{2024NatAs...8..920G}.
A hydrogen-rich atmosphere can escape through the individual or combined action of the planet's own thermal energy, stellar XUV radiation, and tidal forces.
\citet{2024NatAs...8..920G} found that using an upgraded Jeans parameter can determine which mechanism dominates the atmospheric escape.
At the end of planetary formation, planets are in the state of magma ocean and co-evolve with planetary outgassing processes.
% The magma oceans co-evolve with planetary outgassing processes.
Most of the planet's water is slowly released from the magma \citep{2023MNRAS.526.6235M}.
The solidification timescale of the planet's magma ocean is only a few to a few tens of Myr ($1\text{Myr}=10^6\text{years}$) \citep{2016ApJ...829...63S, 2021AsBio..21.1325B}.
After the solidification of the magma ocean and the escape of the hydrogen-rich atmosphere, the atmosphere is dominated by water or \(\text{CO}_2\).
% When the primary atmosphere completely escapes, the secondary atmosphere begins to escape.
% There is evidence for atmospheric escape from the Solar System terrestrial planets.
For planets in solar system, \citet{2019Icar..321..379M} reported net O$^+$ escape rates from Venus of order $10^{24}~\mathrm{s^{-1}}$ based on Venus Express data, while \citet{2023Icar..39314610N} reported ion escape rates from Mars typically of a few $\times10^{24}~\mathrm{s^{-1}}$, increasing to $\sim10^{25}~\mathrm{s^{-1}}$ under stronger solar activity based on Mars Express. The non-thermal escape is currently the main escape mechanism for the atmosphere composed of heavy species although hydrodynamic escape may have occurred in the early stages \citep{2019ApJ...872...99G}.
For exoplanets, the hydrodynamic escape has been discovered on close-in planets. The observations in the ultraviolet wavelength range provide evidence for the escape of hydrogen \citep{2003Natur.422..143V,2013A&A...551A..63B,2014ApJ...786..132K} and oxygen \citep{2015ApJ...804..116B}. \citet{2018ApJ...855L..11O} proposed that the metastable He\,{\sc i} 10833\,\AA\ absorption line provides an ideal new window for probing atmospheric escape from exoplanets, while \citet{2018Natur.557...68S} reported the first detection of this signal in WASP-107b, confirming that it can effectively reveal an extended and escaping planetary atmosphere.
% This phenomenon can be explained by hydrodynamic escape as the drag escape of heavier elements accompanied by H \citep{2023ApJ...953..166X, 2024arXiv241205258S,2024RAA....24f5022X,2025arXiv250318049S}.
The escape of He and O can be explained by hydrodynamic escape as the drag escape of heavier elements accompanied by H \citep{2019ApJ...872...99G, 2023ApJ...953..166X, 2024arXiv241205258S,2024RAA....24f5022X,2025arXiv250318049S}.
It also implies that due to the mass fractionation \citep{2019ApJ...872...99G}, oxygen can remain in the atmosphere when the hydrodynamic flow of H becomes weaker \citep{1987Icar...69..532H}, which further leads to the formation of abiotic oxygen atmospheres on exoplanets \citep{2015AsBio..15..119L, 2016JGRA..121.4718L}.

GJ 486b is orbiting an M dwarf and extremely close to its host star, which has a semi-major axis of only 0.01734 AU (Astronomical unit, 1AU=149,597,871km) \citep{2021Sci...371.1038T, 2022A&A...665A.120C}.
% And, it is also consistent with planet formation predictions that it would accrete a hydrogen-rich atmosphere \citep{2022NatAs...6.1296K}.

Following the planet formation models of \citet{2022NatAs...6.1296K}, GJ 486b should initially have accreted a hydrogen-rich atmosphere.
Based on the retrieval analysis of JWST observations, \citet{2023ApJ...948L..11M} concluded that GJ 486b has a water-rich atmosphere with a water mixing rate higher than 10\%, or that the transmission spectrum is contaminated by water in cool unocculted starspots at the \(2\sigma\) confidence level.
Moreover, \citet{2024ApJ...975L..22W} found that GJ 486b shows no evidence for either a strongly hydrogen-rich atmosphere or a substantial secondary atmosphere, above a small surface-pressure threshold, composed of higher-mean-molecular-weight species such as CO$_2$ and H$_2$O, which hints that the planet should have experienced a violent atmosphere escape.

\citet{2022A&A...665A.120C} employed mass--radius relations together with empirical corrections to analyze the internal structure of GJ 486b \citep{2015A&A...577A..42B,2021Sci...374..330A}.
Their results showed that the bulk composition of GJ 486b is consistent with a rocky, approximately Earth-like composition, while allowing the possibility of a volatile-rich upper layer.
To assess the possible water content of GJ 486b, three interior-structure models with increasing levels of complexity were considered in previous work \citep{2022A&A...665A.120C}.
The simplest model (MR-S) assumes a rocky planet with a thin volatile layer and no compositional constraints.
The MRA-S model introduces stellar abundance ratios (Fe, Mg, Si), which reduce uncertainties but do not account for deep water storage.
The most comprehensive model (MRA-SH) further allows water to dissolve into the molten mantle, reflecting more realistic thermodynamic behavior.
Therefore, in this study, we adopt the MRA-SH result, with a logarithmic water mass fraction of $-4.5^{+1.5}_{-1.3}$, as a representative assumption for the possible present-day water content of GJ 486b.
This corresponds to a water content ranging from 0.021 to 13.04 TO (\(1\,\mathrm{TO}=1.4\times10^{21}\,\mathrm{kg}\)), with a median value of 0.412 TO.

Although water may not definitively exist in GJ 486b, the observations and theoretical studies above make GJ 486b a sutiable example to explore how the initial hydrogen-rich atmosphere affects water escape (remain) and how water inventory evolves with age. Thus, the aim of this study is to use the estimated water content of GJ 486b to infer the initial water content and initial hydrogen-rich atmospheric mass, and to estimate the age of the GJ 486 star in conjunction with planetary formation datasets.
In Section \ref{sec:method}, we describe the calculation methods and parameter settings for hydrogen-rich atmospheres and water escape.
In Section \ref{sec:result}, we present the inversion results for the two dimensions of GJ 486b, as well as the methods and assumptions used to estimate the stellar age, and the stability of the results.
In Section \ref{subsec:effect_magma}, we discuss the reasonable of ignore the magma ocean phase and in Section \ref{subsec:compare}, we compare the GJ 486 star age by our method with previous observation works.
Finally, we summarize our conclusions in Section \ref{sec:con}.

\section{Method} \label{sec:method}

\subsection{Model Descriptions} \label{subsec:description}

In this study, we simulate planetary evolution using VPLanet (2.5.10), an open-source framework designed to model the coupled evolution of stars, planets, and planetary systems \citep{2020PASP..132b4502B}.
VPLanet organizes its computations through modular components that represent different physical processes.
We employ the STELLAR module to simulate stellar evolution and the AtmEsc module to model atmosphere escape processes.

\subsubsection{Stellar Evolution Simulation} \label{subsubsec:stellar}

In VPLanet, the STELLAR module is used to calculate the star parameters (the stellar luminosity and effective temperature) over time by performing a bicubic spline integration across the mass of the stars and time of the evolutionary tracks in \citet{2015A&A...577A..42B}.
Then the empirical broken power-law formula of \citet{2005ApJ...622..680R} is used to track the stellar XUV luminosity as a function of bolometric luminosity.
For M dwarfs, the formula is as follows:

\begin{equation}
\frac{L_{\text{XUV}}}{L_{\text{Bol}}} =
\begin{cases}
f_{\text{sat}}  & \text{if } t \leq t_{\text{sat}} \\
f_{\text{sat}} \left( \frac{t}{t_{\text{sat}}} \right)^{-\beta_{\text{XUV}}} & \text{if } t > t_{\text{sat}}
\end{cases}
\end{equation}

Where \( L_{\text{XUV}} \) is the XUV luminosity of the star.
\( L_{\text{Bol}} \) is the bolometric luminosity which is obtained by interpolating the data of \citet{2015A&A...577A..42B}.
\( t \) is the age from the formation of the star.
\( t_{\text{sat}} \) is the XUV saturation time.
\( \beta_{\text{XUV}} \) is the exponential decay rate of the XUV luminosity.
\( f_{\text{sat}} \) is the XUV saturation fraction, which is the ratio of XUV to bolometric luminosity during the saturation time.
In this work, the last three parameters use the default values in VPLanet.
They are set : \( f_{\text{sat}} = 10^{-3} \), \( t_{\text{sat}} = 10^9 \) years, and \( \beta = 1.23 \) \citep{2020PASP..132b4502B}.

\subsubsection{Atmosphere Escape Simulation} \label{subsubsec:atmospheric}

Based on the evolutionary stage of the planet, atmosphere escape is categorized in our calculations as primary and secondary atmosphere escape.
VPLanet assumes that water does not escape until the hydrogen-rich atmosphere has completely escaped.
We directly use energy-limited formula in AtmEsc module to compute the hydrogen-rich atmosphere escape:

\begin{equation}
\dot{M}_{\text{energy-limited}} = \frac{\epsilon_{\text{XUV}} \pi F_{\text{XUV}} R_p R_{\text{XUV}}^2}{G M_p K_{\text{tide}}} \label{EL}
\end{equation}

Where, \( F_{\text{XUV}} \) is the XUV flux at the orbital distance of the planet.
\( K_\text{tide}  \) is the tidal contribution and we set it to \( K_\text{tide} = 1 \), which means no tidal contribution \citep{2007A&A...472..329E}.
The XUV absorption efficiency, \( \epsilon_{\text{XUV}} \), is typically 1\%-10\%, and we set this value at \( \epsilon_{\text{XUV}} = 0.1 \) \citep{1981Icar...48..150W, 2013MNRAS.430.1247L, 2013ApJ...765...90V}.
\( R_\text{XUV}  \) is the radius at which the planet absorb the XUV radiation.
\( R_\text{XUV}  \) of the primary atmosphere is significantly larger than \( R_p  \).
In VPLanet, the planetary radius relation given by \citet{2012ApJ...761...59L} is used for estimation.

We use the module provided by \citet{2024PSJ.....5..137G} to calculate the water escape, the following is the detailed description and settings.
To estimate the loss of water from terrestrial exoplanets under intense XUV irradiation, we follow the hydrodynamic escape framework developed by \citet{2015AsBio..15..119L} and \citet{2016ApJ...829...63S}.
The reference hydrogen escape flux \(F_{\text{H}}^{\text{ref}}\) is computed in the energy-limited regime by Equation \ref{EL} and the parameter settings are the same as in the previous settings on hydrogen-rich atmosphere escape, except for the XUV absorption efficiency \( \epsilon_{\text{XUV}} \).
We used the XUV absorption efficiencies \( \epsilon_{\text{XUV}} \) for the water vapor atmospheric scenario provided by \citet{2017MNRAS.464.3728B}.
The escape of oxygen occurs through the escape drag of hydrogen.
In the energy-limited escape region, this occurs only when the crossover mass \(m_c\) is greater than the mass of oxygen \(m_O\), or when the incident XUV flux \(F_{\text{XUV}}\) is greater than the critical XUV flux \(F_{\text{XUV}}^\text{Crit}\).
In addition, the drag escape of oxygen is stopped by the diffusion-limited escape region.
In order to calculate the crossover mass \(m_c\), we consider an intermediate variable \(x\) first:

\begin{equation}
x = \left( \frac{m_O}{m_H} - 1 \right)(1 - X_O) \left( \frac{b_{\text{diff}}\, g\, \mu}{k_b T_{\text{flow}}} \right)
 \label{x}
\end{equation}

Where \(m_O\) and \(m_H\) are the masses of oxygen and hydrogen atoms, respectively, \(X_O\) is the mixing rate of oxygen, \(g\) is the gravitational acceleration of the planet's surface, \(\mu\) is the atomic mass per unit, and \(k_b\) is the Boltzmann constant.
The two-body diffusion coefficient for hydrogen and oxygen is \(b_\text{diff}=4.8\times 10^{19}T_\text{flow}^{0.75} \text{m}^{-1}\text{s}^{-1}\).
\(T_\text{flow}\) is the mean thermosphere temperature, and set to 400 K \citep{1996Icar..124..537C, 1987Icar...69..532H}.
By introducing \(x\), we can express the crossover mass in the form of a segmented function \citep{2015AsBio..15..119L}:

\begin{equation}
m_c =
\begin{cases}
\mu + \dfrac{1}{1-X_O}\dfrac{k_bT_\text{flow}F_{\text{H}}^{\text{ref}}}{b_\text{diff}g} &
F_{\text{H}}^{\text{ref}} < x \\[10pt]
\mu\times\dfrac{1+(\frac{X_O}{1-X_O})(\frac{m_O}{m_H})^2}{1+(\frac{X_O}{1-X_O})(\frac{m_O}{m_H})} + \dfrac{1}{1+X_O(\frac{m_O}{m_H}-1)}\dfrac{k_bT_\text{flow}F_{\text{H}}^{\text{ref}}}{b_\text{diff}g}&
F_{\text{H}}^{\text{ref}} > x
\end{cases}
\label{mc}
\end{equation}

The critical flux of incident XUV is defined by \citet{2016ApJ...829...63S}:

\begin{equation}
F_{\text{XUV}}^\text{Crit}=
\frac{4b_\text{diff}\nu_H^2}{\varepsilon_\text{XUV}k_bT_\text{flow}R_p}(\frac{m_O}{m_H}-1)(1-X_O)
\label{XUVc}
\end{equation}

Where, the thermal velocity of hydrogen is \(\nu_H=\sqrt{\frac{2k_bT_\text{flow}}{m_H}}\).
The drag escape of oxygen occurs only when the incident XUV flux is greater than the critical flux and the crossover mass is greater than the mass of oxygen, at this time the escape rate of hydrogen is corrected to:

\begin{equation}
F_H= F_{\text{H}}^{\text{ref}}\times
(1+\frac{X_O}{1-X_O}\frac{m_O}{m_H}\frac{m_c-m_O}{m_c-m_H})^{-1}
\label{FH1}
\end{equation}

When the mixing ratio of oxygen in the atmosphere is greater than that of water (\(X_O>X_{H_2O}\)), the diffusion-limited escape region is entered, at this time the escape rate of hydrogen is \citep{2015AsBio..15..119L}:

\begin{equation}
F_H = b_\text{diff}g\mu\frac{(m_O-m_H)(1-X_O)}{m_Hk_bT_\text{flow}}
\label{FH2}
\end{equation}

With the escape rate of hydrogen, we can give the total atmospheric mass loss \citep{2015AsBio..15..119L}:

\begin{equation}
\dot M = 4\mu\pi R_p^2 F_H
\label{Mdot}
\end{equation}

Furthermore, the lifetimes of protoplanetary disks of low-mass stars are in the range of 4.2-5.8 Myr, as reported in the observations of \citet{2014A&A...561A..54R}.
Therefore, we set 5 Myr as the lifetime of the protoplanetary disk of GJ 486.
This means that planets begin to evolve when the age of the star reaches 5 Myr, which is also consistent with previous works \citep{2016ApJ...829...63S, 2021AsBio..21.1325B, 2024PSJ.....5..137G}.

\subsection{Input Description} \label{subsec:input}

We use the stellar and planetary parameters determined by \citet{2022A&A...665A.120C} and \citet{2024A&A...689A..48D} as inputs into the simulations. The stellar mass is set to 0.31 M$_{\odot}$. The planetary mass is set to 3.0 M$_{\oplus}$ and the orbit semi-major axis is set to \(0.017\) AU \citep{2022A&A...665A.120C}.
\citet{2024PSJ.....5..137G} reported the presence of outliers larger than 4 standard deviations in the sample of stellar masses simulated by VPLanet.
They suggested that this was likely an artifact of the machine learning approach implemented by \citet{2021RNAAS...5..122B}.
However, since our simulations do not account for tidal effects, and possibly because of the use of a higher version of VPLanet (2.5.10) this phenomenon is not detected in our simulations.

In our task, the initial water content of the planet and the mass of the primary atmosphere are inversion variables.
The age of the star is a parameter of the model and does not participate in the inversion, but we generate its gridded data for analyzing its effects.
Previous studies have been very inconsistent in determining the period and age of GJ 486 due to its low luminosity \citep{2022A&A...665A.120C}.
The ASAS-SN and SuperWASP (North and South combined) light curves indicate a rotation period of 130 days for GJ 486, which corresponds to the age of $7.8 \pm 4.8$ Gyr ($1\text{Gyr}=10^9\text{years}$) in the age-rotation relationship for M2.5-6.5 dwarfs with rotation period more than 24 days \citep{2021Sci...371.1038T, 2023ApJ...954L..50E, 2024A&A...689A..48D}. However, this is not in agreement with the result of \citet{2022A&A...665A.120C}.
% They suggest the rotation period of GJ 486 is $49.9\pm5.5$ days, corresponding to the stellar age range of $3.7^{+5.1}_{-\text{Univ}}$ Gyr \citep{2024A&A...689A..48D}.
They suggest the rotation period of GJ 486 is $49.9\pm5.5$ days, corresponding to the stellar age range of $3.7^{+5.1}_{-3.7}$ Gyr \citep{2024A&A...689A..48D}.
Based on the activity-age relationship, the panchromatic spectrum from 5 to 1700 {\AA} of GJ 486, has an age range of $10.5^{+\text{Univ}}_{-3.9}$ Gyr \citep{2024A&A...689A..48D}.
The "Univ" means uncertainty extends to the age of the universe.
Therefore, in this study we consider the age of the star as a free parameter.

In simulations, the initial water content of the planets ranges from 0 to 370 TO in steps of 0.25 TO.
The range of planetary primary atmosphere mass is \(0-0.16 M_\oplus\) in steps of \(0.004 M_\oplus\).
Although the interpolation tables of stellar ages and physical parameters provided by \citet{2015A&A...577A..42B} only extend to 10 Gyr, we adopt the last available value to represent all subsequent ages because low-mass stars like GJ 486 barely evolve.
Thus, the age range of the stars is 0-13 Gyr with a step size of 50 Myr.
For the simulations, we generate a total of 60,721 data points, each characterized by five parameters:

the initial water content (\(M_w\)), the initial mass of the primary atmosphere (\(M_\text{H/He}\)), the stellar age (\(t\)), the final water content of the planet after evolution (\(M_w'\)), and the final mass of the hydrogen-rich atmosphere (\(M_\text{H/He}'\)).

\section{Result} \label{sec:result}

\citet{2024ApJ...975L..22W} considered that GJ 486b not only doesn't have a significantly hydrogen-rich atmosphere but also doesn't have a substantial secondary atmosphere (above some small surface pressure) composed of higher mean molecular weight molecules (e.g., CO2, H2O, etc.).
This indicates that the planet currently has no atmosphere.
It is relatively clear that early hydrogen can completely escape. However, the water that existed in the atmosphere in the early phase may have already settled on the surface or interior of the planet so that the planet currently contains water. Therefore, we use the conditions of planets having water and no hydrogen in their atmosphere to limit their early states. The current water content of GJ 486b ranges from 0.021 to 13.04 TO, with a median value of 0.412 TO \citep{2022A&A...665A.120C}.

%We use these two conditions to invert the initial water content and primary hydrogen-rich atmospheric mass at different stellar ages.
%However, our model cannot distinguish between liquid water and atmospheric water, and we can only refer to the part of the results that does not %contain hydrogen-rich atmosphere as the inversion target.

\subsection{Water Content Inversion at Previous Age} \label{sec:r1}

To quantify the constraints on the initial water content of GJ~486b, we perform a grid of evolutionary simulations spanning a wide range of initial conditions.
Specifically, the initial hydrogen-rich atmospheric mass is varied between 0 and 0.16 \(M_\oplus\), while the stellar age is sampled from 0 to 13 Gyr.
For each model, we record the combinations of initial water content and hydrogen-rich atmosphere that can reproduce a given present-day water content.

Based on current observational constraints, we focus on three representative values of the present water content, 0.021, 0.412, and 13.04~TO.
These values bracket the plausible range inferred for GJ~486b and serve to illustrate how the inversion depends on the assumed present-day state.

\begin{figure*}
    \centering
    \includegraphics[width=\textwidth]{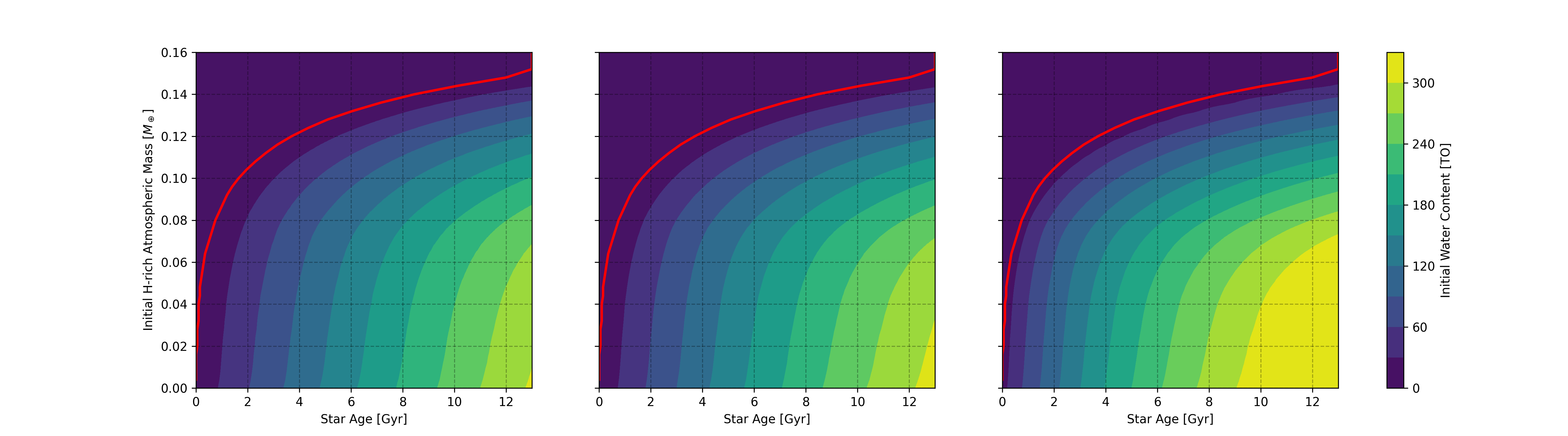}
    \caption{
Inversion results for current water content of 0.021, 0.412 and 13.04 TO.
The three plots from left to right represent the cases when the current water content is 0.021, 0.412 and 13.04 TO, respectively.
Each plot represents the initial water content required to reach the current water content for different initial hydrogen-rich atmospheric masses and stellar ages.
The red line indicates the cutoff where the current hydrogen-rich atmospheric mass is zero. In the upper region of the red line, the planet still has a hydrogen-rich atmosphere at the end of its evolution.
}
    \label{fig1-1}
\end{figure*}

Figure~\ref{fig1-1} illustrates the inversion results for the three present-day water contents.
In each panel, the red curve indicates the cutoff at which the hydrogen-rich atmosphere is completely removed by the end of the evolution.
Models located above this boundary retain a hydrogen-rich atmosphere at the present epoch.
Such outcomes are inconsistent with current observational constraints on GJ~486b \citep{2024ApJ...975L..22W} and are therefore excluded from further consideration.
Several systematic trends are apparent.
At fixed stellar age, a larger initial hydrogen-rich atmospheric mass reduces the initial water content required to match the present-day value.
This reflects the shielding effect of the hydrogen-rich envelope, which delays the onset of water loss.
Conversely, for a fixed initial hydrogen-rich atmospheric mass, the required initial water content increases with stellar age, owing to the longer cumulative exposure to stellar irradiation and atmospheric escape.

\begin{figure}
    \centering
    \includegraphics[width=0.5\linewidth]{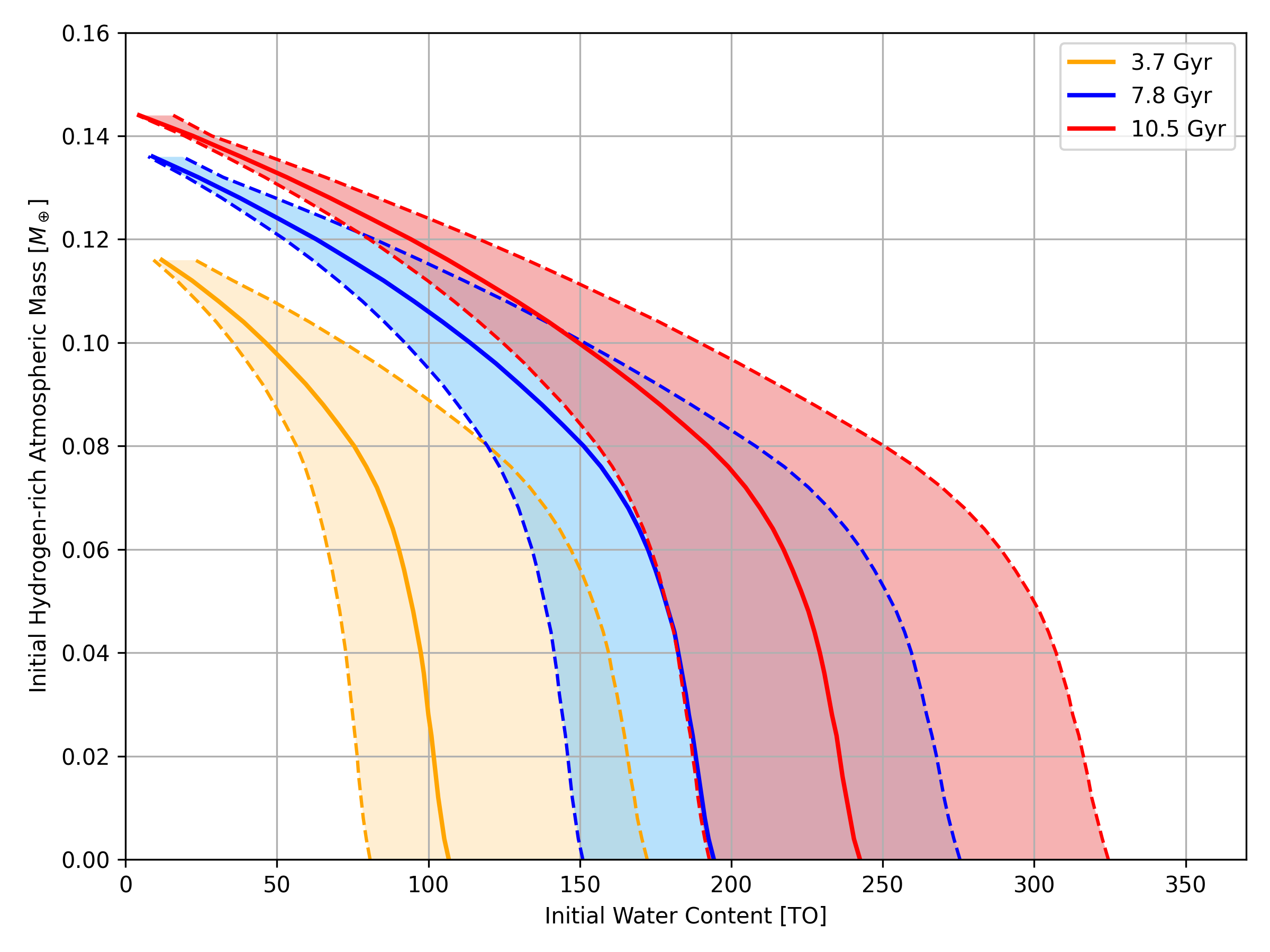}
\caption{
Initial water content inversion curves at fixed stellar ages of 3.7, 7.8 and 10.5~Gyr.
The X-axis indicates the initial water content in units of TO, while the Y-axis shows the initial hydrogen-rich atmospheric mass in units of $M_\oplus$.
The orange, blue, and red curves correspond to the inversion results for the median present-day water content of 0.412~TO at 3.7, 7.8, and 10.5~Gyr, respectively.
For each age slice, the surrounding envelope indicates the allowed inversion range when the present-day water content varies from 0.021 to 13.04~TO.
The corresponding dashed lines represent the boundaries of the inversion targets at 0.021 and 13.04~TO, respectively.
}
    \label{fig3_res}
\end{figure}

To more clearly illustrate the relationship between the initial water content ($M_\mathrm{w}$) and the initial hydrogen-rich atmospheric mass ($M_{\mathrm{H/He}}$), we fix the stellar age at 3.7, 7.8, and 10.5~Gyr reported in previous studies \citep{2024A&A...689A..48D}, and extract the corresponding inversion curves from the MRA-SH model.
For each age, the solid curve in Figure~\ref{fig3_res} shows the locus of $(M_\mathrm{w}, M_{\mathrm{H/He}})$ pairs that reproduce the median present-day water content of 0.412~TO, while the envelope brackets the solutions obtained when the present-day water content is varied from 0.021 to 13.04~TO.

Figure~\ref{fig3_res} highlights a pronounced degeneracy between $M_\mathrm{w}$ and $M_{\mathrm{H/He}}$.
In the limit of $M_{\mathrm{H/He}} = 0$, matching the same present water content requires a substantially larger initial water reservoir, and the required $M_\mathrm{w}$ increases systematically with stellar age.
Quantitatively, for the median present-day water content of 0.412~TO, the $M_\mathrm{w}$ increases from 107~TO (3.7~Gyr) to 194~TO (7.8~Gyr) and 243~TO (10.5~Gyr) when $M_{\mathrm{H/He}}=0$.
The corresponding allowed ranges (envelopes) also shift to higher values, from 81--172~TO at 3.7~Gyr to 151--276~TO at 7.8~Gyr and 193--325~TO at 10.5~Gyr.
This trend reflects the cumulative effect of long-term atmospheric escape, namely, older planets experience a longer loss history.
Thus, a larger primordial water budget is necessary when no protective hydrogen-rich atmosphere is present.

% Once a hydrogen-rich atmosphere is included, the inferred initial water content drops rapidly, indicating that even a modest primordial envelope can strongly reduce the required $M_\mathrm{w}$.
% For example, at $M_{\mathrm{H/He}}=0.06~M_\oplus$, the $M_\mathrm{w}$ decreases from 107 to 90~TO (a reduction of 17~TO) at 3.7~Gyr, from 194 to 173~TO (a reduction of 21~TO) at 7.8~Gyr, and from 243 to 217~TO (a reduction of 26~TO) at 10.5~Gyr.
% The reduction becomes even more pronounced for $M_{\mathrm{H/He}}=0.10~M_\oplus$, where $M_\mathrm{w}$ drops to 46~TO (3.7~Gyr), 114~TO (7.8~Gyr), and 150~TO (10.5~Gyr), corresponding to decreases of 61~TO, 80~TO, and 93~TO relative to the $M_{\mathrm{H/He}}=0$ solutions, respectively.
% In this regime, the hydrogen-rich atmosphere effectively buffers the water reservoir by delaying its exposure to escape-driven depletion, so that the inversion relation becomes steep and increasingly controlled by $M_{\mathrm{H/He}}$.

Once a hydrogen-rich atmosphere is included, the inferred initial water content decreases rapidly, indicating that even a modest primordial envelope can substantially reduce the required $M_\mathrm{w}$ (Figure~\ref{fig3_res} and Table~\ref{tab:mw_reduction}).
For example, at $M_{\mathrm{H/He}}=0.06~M_\oplus$, the inferred $M_\mathrm{w}$ is reduced from 107 to 90~TO at 3.7~Gyr, from 194 to 173~TO at 7.8~Gyr, and from 243 to 217~TO at 10.5~Gyr, corresponding to reductions of 17, 21, and 26~TO, respectively.
The effect becomes much prominent at $M_{\mathrm{H/He}}=0.10~M_\oplus$, where $M_\mathrm{w}$ decreases to 46~TO, 114~TO, and 150~TO at 3.7, 7.8, and 10.5~Gyr, respectively, corresponding to reductions of 61, 80, and 93~TO relative to the $M_{\mathrm{H/He}}=0$ solutions.
As summarized in Table~\ref{tab:mw_reduction}, the reduction becomes systematically larger both with increasing system age and with increasing hydrogen-rich atmosphere mass.
% This trend indicates that the hydrogen-rich atmosphere effectively buffers the water reservoir by delaying its exposure to escape-driven depletion, so that the inversion relation becomes steeper and increasingly controlled by $M_{\mathrm{H/He}}$.
In this regime, the hydrogen-rich atmosphere effectively buffers the water reservoir by delaying its exposure to escape-driven depletion, so that the inversion relation becomes steep and increasingly controlled by $M_{\mathrm{H/He}}$.

% \begin{table}[htbp]
% \centering
% \caption{Representative situation of reductions in the inferred initial water content $M_\mathrm{w}$ after including a hydrogen-rich atmosphere.}
% \label{tab:mw_reduction}
% \begin{tabular}{cccc}
% \hline
% Age (Gyr) & $M_{\mathrm{H/He}}$ ($M_\oplus$) & Inferred $M_\mathrm{w}$(TO)  &  Reduction (TO) \\
% \hline
% 3.7  & 0.00 & 107 & --- \\
% 7.8  & 0.00 & 194 & --- \\
% 10.5 & 0.00 & 243 & --- \\
% 3.7  & 0.06 & 90  & 17 \\
% 7.8  & 0.06 & 173 & 21 \\
% 10.5 & 0.06 & 217 & 26 \\
% 3.7  & 0.10 & 46  & 61 \\
% 7.8  & 0.10 & 114 & 80 \\
% 10.5 & 0.10 & 150 & 93 \\
% \hline
% \end{tabular}
% \end{table}

\begin{deluxetable}{cccc}
\tablecaption{Reductions in the inferred initial water content $M_\mathrm{w}$ after including a hydrogen-rich atmosphere.\label{tab:mw_reduction}}
\tablewidth{0pt}
% \tablewidth{\textwidth}
\tablehead{
\colhead{Age} & \colhead{$M_{\mathrm{H/He}}$} & \colhead{$M_\mathrm{w}$} & \colhead{Reduction} \\
\colhead{(Gyr)} & \colhead{($M_\oplus$)} & \colhead{(TO)} & \colhead{(TO)}
}
\startdata
3.7  & 0.00 & 107 & \nodata \\
     & 0.02 & 102 & 5 \\
     & 0.04 & 98  & 9 \\
     & 0.06 & 90  & 17 \\
     & 0.08 & 76  & 31 \\
     & 0.10 & 46  & 61 \\
\hline
7.8  & 0.00 & 194 & \nodata \\
     & 0.02 & 188 & 6 \\
     & 0.04 & 183 & 11 \\
     & 0.06 & 173 & 21 \\
     & 0.08 & 151 & 43 \\
     & 0.10 & 114 & 80 \\
     & 0.12 & 63  & 131 \\
\hline
10.5 & 0.00 & 243 & \nodata \\
     & 0.02 & 236 & 7 \\
     & 0.04 & 229 & 14 \\
     & 0.06 & 217 & 26 \\
     & 0.08 & 192 & 51 \\
     & 0.10 & 150 & 93 \\
     & 0.12 & 94  & 149 \\
     & 0.14 & 22  & 221 \\
\enddata
\tablecomments{
The reduction is defined with respect to the corresponding case with $M_{\mathrm{H/He}}=0.00\,M_\oplus$ at the same age.
Hence, the reduction for $M_{\mathrm{H/He}}=0.00\,M_\oplus$ itself is denoted by \nodata.}
\end{deluxetable}

\begin{figure}
    \centering
    \includegraphics[width=\textwidth]{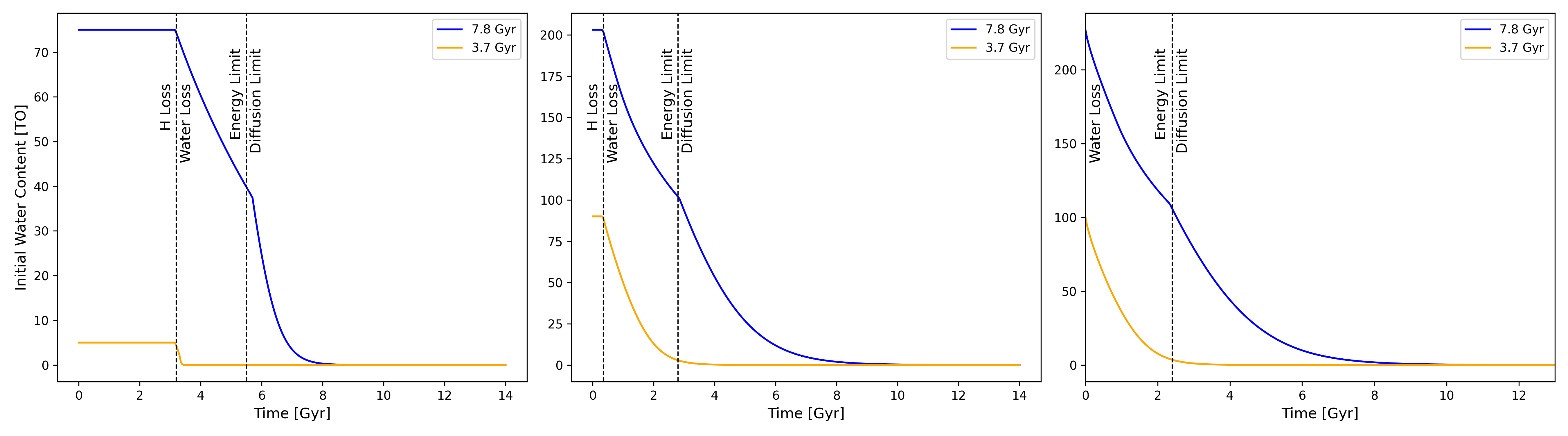}
    \caption{
Evolution of the planetary water content for GJ~486b under different assumptions on stellar age and initial hydrogen-rich atmospheric mass.
Blue and orange curves correspond to stellar ages of 7.8~Gyr and 3.7~Gyr, respectively.
The left, middle, and right panels show cases with initial hydrogen-rich atmospheric masses of 0.116, 0.060, and 0.000 \(M_\oplus\), respectively.
In the left and middle panels, the first dashed vertical line marks the end of hydrogen-rich atmospheric escape and the onset of water loss, while the second dashed line indicates the transition from energy-limited to diffusion-limited escape.
In the right panel, water escape begins immediately because no primordial hydrogen-rich atmosphere is present.
}
    \label{fig0}
\end{figure}

We further show how our model constrains the initial water content of GJ~486b through inversion, and how this constraint depends on the age of the host star. Following the discussion in Section~\ref{subsec:input}, we consider two representative stellar ages, 3.7~Gyr and 7.8~Gyr. Figure~\ref{fig0} shows the evolution of the planetary water content for these two age assumptions.
The left, middle, and right panels correspond to initial hydrogen-rich atmospheric masses of 0.116, 0.060, and 0.000 \(M_\oplus\), respectively.
For the cases with an initial hydrogen envelope, two dashed vertical lines mark key transitions: the end of hydrogen-rich atmospheric escape and the subsequent transition from energy-limited to diffusion-limited escape.
In the case without a hydrogen-rich atmosphere (right panel), water escape begins immediately.
A clear separation between hydrogen-rich atmospheric escape and water loss is evident in other cases.
During the hydrogen escape phase, the water content remains essentially constant (left and middle panels of Figure \ref{fig0}).
As the host star evolves and its XUV luminosity decreases, the hydrogen escape rate gradually declines.
Once it falls below a critical threshold, hydrogen can no longer effectively carry oxygen away.
At this point, water loss transitions from the energy-limited regime to the diffusion-limited regime, as indicated by the second dashed line in the left and middle panels.
In the diffusion-limited phase, water escape becomes significantly less efficient and proceeds on much longer timescales.
Eventually, as the stellar XUV irradiation weakens further, water escape effectively ceases and the remaining water content approaches a constant value. This final water inventory corresponds to current state of the planet and provides the reference for this inversion.

The dependence on stellar age is also apparent.
For the older system (7.8~Gyr), the longer cumulative exposure to stellar irradiation results in more substantial water loss, requiring a larger initial water content to reproduce the same final state.
In contrast, for the younger system (3.7~Gyr), the shorter evolutionary timescale limits water loss, allowing a smaller initial water reservoir to remain consistent with the observations.

\subsection{Constraints on the Age of GJ 486 by Planetary Formation} \label{age}

% \subsubsection{\textcolor{red}{Analysis of Planetary Formation Dataset}} \label{subsec:DatasetAnalysis}

\subsubsection{Analysis of Planetary Formation Dataset}
\label{subsec:DatasetAnalysis}
The above results indicate that the initial hydrogen rich atmosphere and water inventory can fit the current water content over a wide range attributed to the uncertainty of age. Therefore, determining the age of planets becomes important. In this section, we propose a method for estimating age based on the dataset of planetary formation. We introduce the enriched planet formation data for 0.3 M$_{\odot}$ stars provided by \citet{2022NatAs...6.1296K}.
Here we focus on describing our approach to using this dataset and the constraints it places on our inversion results.
Given that the dataset presents planet formation as a scatter distribution, directly analyzing the specific case of a semi-major axis \(a_0 = 0.01734\) AU and a planetary mass \(M_{p0} = 3M_{\oplus}\) is not feasible.
Instead, we explore the phase space \((a, M_p)\) of semi-major axes and planetary masses, considering planets within the regions \(a_0 \pm \Delta a\) and \(M_{p0} \pm \Delta M_p\) as possible scenarios for GJ 486b.

\begin{figure}
    \centering
    \includegraphics[width=0.5\linewidth]{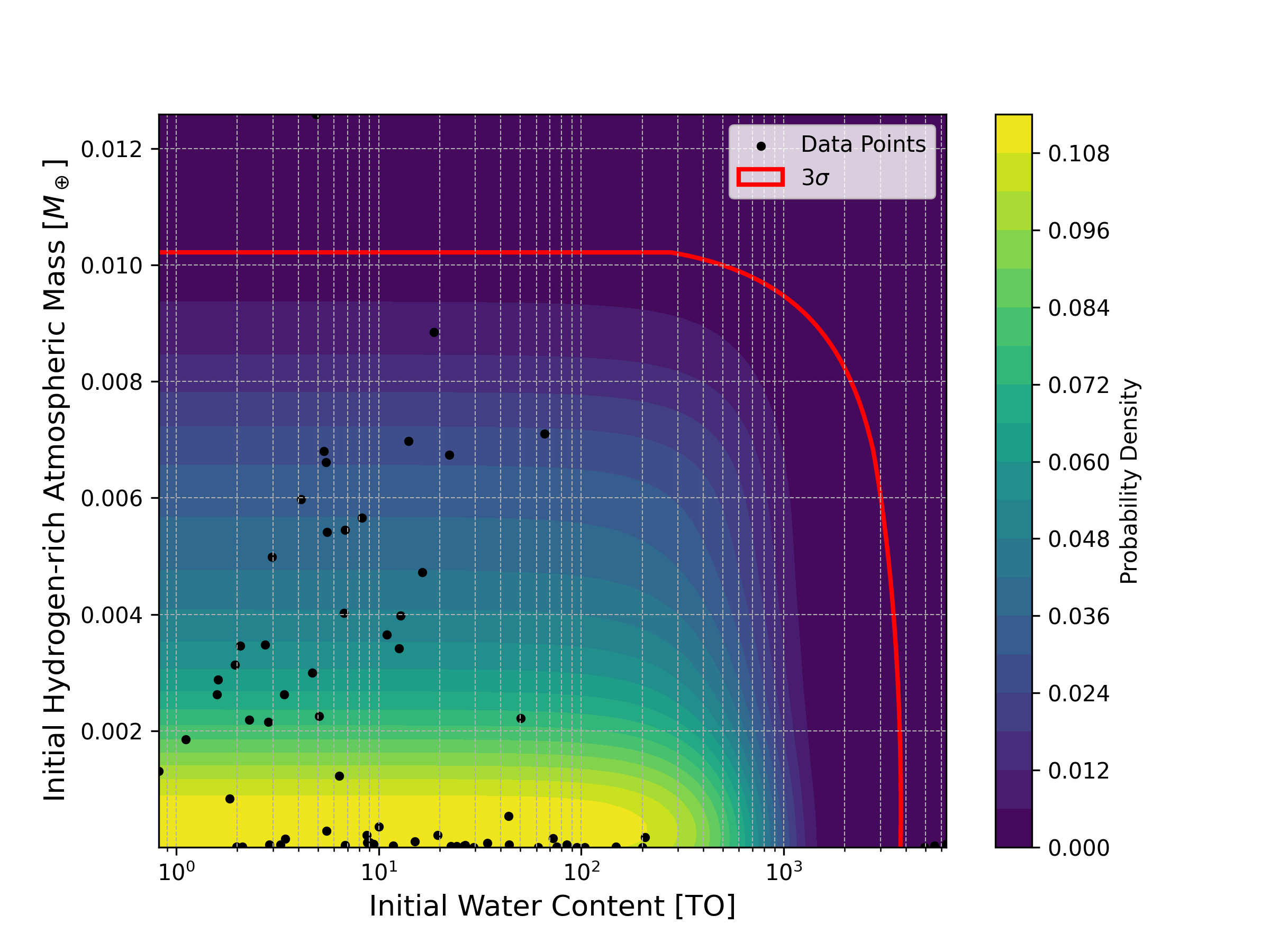}
    \caption
{
Black points are the sample of the planetary formation dataset provided by \citet{2022NatAs...6.1296K} in the phase space of \(\Delta a=0.005\)AU, \(\Delta M_p=0.5M_{\oplus}\).
Two-dimensional probability density distribution of the initial water content and hydrogen-rich atmospheric mass for GJ 486b analogs, constructed using Gaussian kernel density estimation (KDE).
The water content axis is sampled logarithmically and the atmospheric mass axis is sampled linearly to properly account for the dynamic range of the variables.
The red overlaid ellipse represents the \(3\sigma\) confidence region derived from the covariance matrix, indicating the principal dispersion of the sample distribution.
}
    \label{appfig1}
\end{figure}

Figure \ref{appfig1} illustrates the sample of the planetary formation dataset in the phase space of \(\Delta a=0.005\)AU, \(\Delta M_p=0.5M_{\oplus}\).
To statistically characterize the physical properties (specifically, the initial water content and the initial hydrogen-rich atmospheric mass) within this sample, we applied Gaussian kernel density estimation (KDE) to the selected points.
The KDE was performed on a two-dimensional grid, where the water content axis was sampled logarithmically to properly account for the wide dynamic range, while the atmospheric mass axis was sampled linearly.
Figure \ref{appfig1} also presents the resulting two-dimensional probability density distribution.
From this distribution, we computed the statistical mean vector and the covariance matrix of the sample.
Based on the covariance matrix, we derived a \(3\sigma\) confidence ellipse, corresponding to the principal axes of the data dispersion.
As a result, we find that the hydrogen-rich atmospheric mass of a planet similar to GJ 486b would not exceed 0.01, depending on the constraints of the planetary formation dataset.

% \subsubsection{\textcolor{red}{Statistical Results of GJ 486 Age}}

\subsubsection{Statistical Results of GJ 486 Age}  \label{GJ_486_Age}

In the section \ref{sec:r1}, we discussed the inversion results when the stellar age is a free parameter of 0-13 Gyr.
However the likelihood between each stellar age is not equiprobable based on our analysis of the planet formation dataset (see Section \ref{subsec:DatasetAnalysis}).
We consider two limiting cases to motivate our definition of the probability distribution over stellar ages.
On the one hand, if the region corresponding to the inversion result of a particular stellar age contains no points from the planetary formation dataset, then it is unlikely that this stellar age represents the true scenario.
On the other hand, if the region associated with a given stellar age covers nearly the entire sample space, then although it may contain a large number of planetary formation outcomes, the result lacks specificity and is therefore not representative.

In Appendix \ref{app1}, we designed a method to estimate stellar ages by combining water content inversion results with planetary formation datasets, and applied it to the GJ 486 system.
The resulting value, $\mu = 2.90\, \mathrm{Gyr}$, is indicated by the red dashed line in Figure \ref{fig3}.
Thus, we conclude that GJ 486 is likely to be a star with an expected age of $2.90^{+2.47}_{-2.27}~\mathrm{Gyr}$ and a \(3\sigma\) range of 0 to 10.85 Gyr based on our inversion and the planetary formation dataset by \citet{2022NatAs...6.1296K} with the phase space of \(\Delta a=0.005\)AU, \(\Delta M_p=0.5M_{\oplus}\).
Figure \ref{fig5} shows our estimates of the age of the star when different phase spaces of the planetary formation dataset are selected.
We find that the mean value of the stellar age \(\mu\) is around 3 Gyr and hardly changes with the selection of phase space.
The right boundary of the \(3\sigma\) range is in the range of 10-12 Gyr.
Therefore, we conclude that the range of phase space does not significantly affect the results of the stellar ages, which average around 3 Gyr.

% To evaluate the age of the GJ~486 system, we have re-determined the age from their kinematics.
To independently validate our method-based estimate, we have re-determined the age from their kinematics.
Based on the astrometry and radial velocity data from \emph{Gaia} DR3, we calculated their Space Position and Galactic velocity with the same procedure decribed in Section 2.1 of \citet{2021ApJ...909..115C}.
The GJ~486 lies very close to the Galactic plane (with a vertical height of only $0.0327~\mathrm{kpc}$) and the total velocity $V_{\rm tot}$ is $40.1~\mathrm{km~s^{-1}}$.
Its vertical angular momentum $L_{z}$ and orbital actions $(J_{R}, J_{z})$ are $2046.8~\mathrm{kpc~km~s^{-1}}$ and $(26.11~\mathrm{kpc~km~s^{-1}}, 0.63~\mathrm{kpc~km~s^{-1}})$, respectively.
These characteristics indicate that it is a typical thin-disk star (with $V_{\rm tot} < 70~\mathrm{km~s^{-1}}$, $L_{z} \sim 2000~\mathrm{kpc~km~s^{-1}}$, $J_{R} < 50~\mathrm{kpc~km~s^{-1}}$, and $J_{z} < 10~\mathrm{kpc~km~s^{-1}}$ \citep{2014A&A...562A..71B,2019A&A...631A..47K,2021AJ....162..100C}).

With the refined kinematic method and updated typical characteristics \citep{2021ApJ...909..115C,2023AJ....166..243Y}, we calculate its probabilities belonging to thick disk over thin disk and the resulted TD/D is 0.0403.
To obtain their kinematic age, we selected planet host stars with similar TD/D, angular momentum and orbital actions (i.e., the relative difference is no greater than 10\%) and then calculate their kinematic age from the velocity dispersion with the refined age-velocity dispersion from \citet{2021ApJ...909..115C}.
The kinematic age for GJ 486 is $3.15^{+0.48}_{-0.35}~\mathrm{Gyr}$, which is consistent with that from the planetary water content inversion.

\subsubsection{Water Content Inversion Results in Our Age Result}
\label{subsec:inv_at_age}

In Section~\ref{GJ_486_Age}, we inferred that GJ~486 most likely has an expected stellar age of
$2.90^{+2.47}_{-2.27}~\mathrm{Gyr}$, with a $3\sigma$ upper bound of 10.85~Gyr.
Following the same inversion procedure described in Section~\ref{sec:r1}, we fix the stellar age to
$t = 2.9~\mathrm{Gyr}$ and extract the locus of initial conditions that reproduce the current water content (0.412~TO) predicted by the MRA-SH model.

\begin{deluxetable}{cccc}
\tablecaption{Reductions in the inferred initial water content $M_\mathrm{w}$ after including a hydrogen-rich atmosphere for stellar age at 2.9~Gyr.\label{tab:mw_reduction29}}
\tablewidth{0pt}
% \tablewidth{\textwidth}
\tablehead{
\colhead{Age} & \colhead{$M_{\mathrm{H/He}}$} & \colhead{$M_\mathrm{w}$} & \colhead{Reduction} \\
\colhead{(Gyr)} & \colhead{($M_\oplus$)} & \colhead{(TO)} & \colhead{(TO)}
}
\startdata
2.9  & 0.00 & 88 & \nodata \\
     & 0.02 & 83 & 5 \\
     & 0.04 & 79 & 9 \\
     & 0.06 & 72 & 16 \\
     & 0.08 & 58 & 30 \\
     & 0.10 & 30 & 58 \\
\enddata
\tablecomments{
This table is identical in format to Table~\ref{tab:mw_reduction}, but presents the results for stellar age at 2.9~Gyr.}
\end{deluxetable}

\begin{figure}
    \centering
    \includegraphics[width=0.55\linewidth]{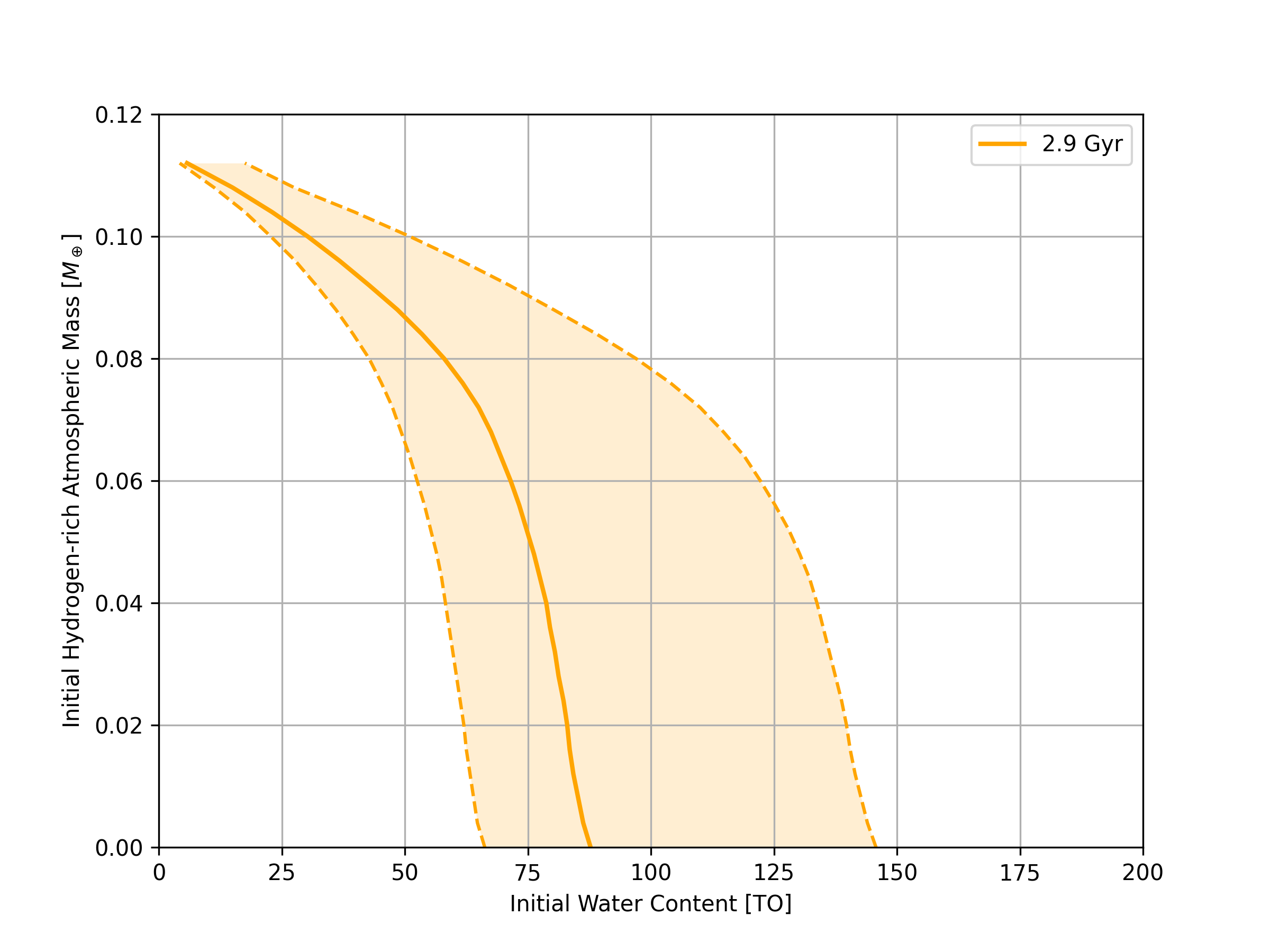}
    \caption{
Initial water content inversion results evaluated at the expected stellar age of $2.9~\mathrm{Gyr}$.
The x-axis indicates the initial water content $M_\mathrm{w}$ in units of TO, and the y-axis indicates the initial hydrogen-rich atmospheric mass $M_{\mathrm{H/He}}$ in units of $M_\oplus$.
The orange curve shows the inversion curve that reproduces the representative present-day water content of 0.412~TO predicted by the MRA-SH model.
The surrounding envelope indicates the corresponding inversion range when the present-day water content is varied from 0.021 to 13.04~TO.
The dashed lines represent the boundaries of the inversion targets at 0.021 and 13.04~TO, respectively.
}
    \label{fig4_res}
\end{figure}

Figure~\ref{fig4_res} shows the inversion result at stellar age $t = 2.9~\mathrm{Gyr}$.
The figure clearly illustrates the degeneracy between the initial water reservoir and the primordial
hydrogen-rich atmospheric mass.
In the limiting case of $M_{\mathrm{H/He}}= 0$, matching the present-day water content requires an initial water content the range of 66 to 146~TO and the middle of 88~TO.
% In contrast, when a hydrogen-rich atmosphere is present, the inferred $M_\mathrm{w}$ decreases rapidly, indicating that even a moderate atmosphere can substantially reduce the initial water requirement.
% This behaviour is consistent with the role of the hydrogen-rich atmosphere as an effective buffer that delays the exposure of the water content to depletion.
Conversely, a hydrogen-rich atmosphere can significantly reduce the initial water content for the inversion results.
Table~\ref{tab:mw_reduction29} shows the detailed water content reduction after including a hydrogen-rich atmosphere.
Quantitatively, the water content decreases from $M_{\mathrm{w}}\simeq 88$~TO at $M_{\mathrm{H/He}}=0$ (with an allowed range of 66--146~TO) to $M_{\mathrm{w}}\simeq 83$~TO at $M_{\mathrm{H/He}}=0.02~M_\oplus$ (62--140~TO), and further to $M_{\mathrm{w}}\simeq 72$~TO at $M_{\mathrm{H/He}}=0.06~M_\oplus$ (53--122~TO).
The hydrogen-rich atmosphere of $0.10~M_\oplus$ reduces the requirement to $M_{\mathrm{w}}\simeq 30$~TO (23--51~TO), a reduction of about 58~TO compared to no hydrogen-rich atmosphere.
A large $M_{\mathrm{H/He}}$ prolongs the epoch during which mass loss is dominated by hydrogen-rich atmosphere, so significant water depletion begins later when the stellar XUV irradiation is weaker, thereby reducing the integrated water loss and decreasing the inferred initial $M_{\mathrm{w}}$.

\section{Discussion} \label{sec:dis}

\subsection{Benchmarking Our Age Inference on TRAPPIST-1 e and f}
\label{subsec:TR1test}

The TRAPPIST-1 system has a clear and widely used method for estimating water content \citep{2021PSJ.....2....1A,2024PSJ.....5..137G}.
Given this combination of rich observational characterisation and multi method age constraints, TRAPPIST-1 stands out as the most suitable system in our sample with a well established reference age.
Therefore, in this subsection, we select TRAPPIST-1 (TR-1) as the primary validation target for our age inference method.

% \textcolor{red}{
% To validate our method, we apply it to a benchmark system with an extensively studied age, TRAPPIST-1 (TR-1).
We use TR-1e and TR-1f as the benchmark target because TR-1b, c, and d are almost dry, while TR-1g and h have large semi-major axes, resulting in only a few planetary formation samples around their vicinity.
The adopted planetary properties are as follows from \citet{2021PSJ.....2....1A}.
For TR-1e, the mass is 0.693 $M_{\oplus}$, the semi-major axis is 0.02925 AU, and the present-day water content ranges from 42.18 to 138.6 TO.
For TR-1f, the mass is 1.040 $M_{\oplus}$, the semi-major axis is 0.03849 AU, and the present-day water content ranges from 149.22 to 284.87 TO.
Following \citet{2021PSJ.....2....1A}, both TR-1e and TR-1f are expected to host secondary atmospheres rather than hydrogen-rich primordial ones.
Accordingly, in our inversion we set the current hydrogen-rich atmospheric mass to zero.
% }

\begin{figure}
    \centering
    \includegraphics[width=1.0\linewidth]{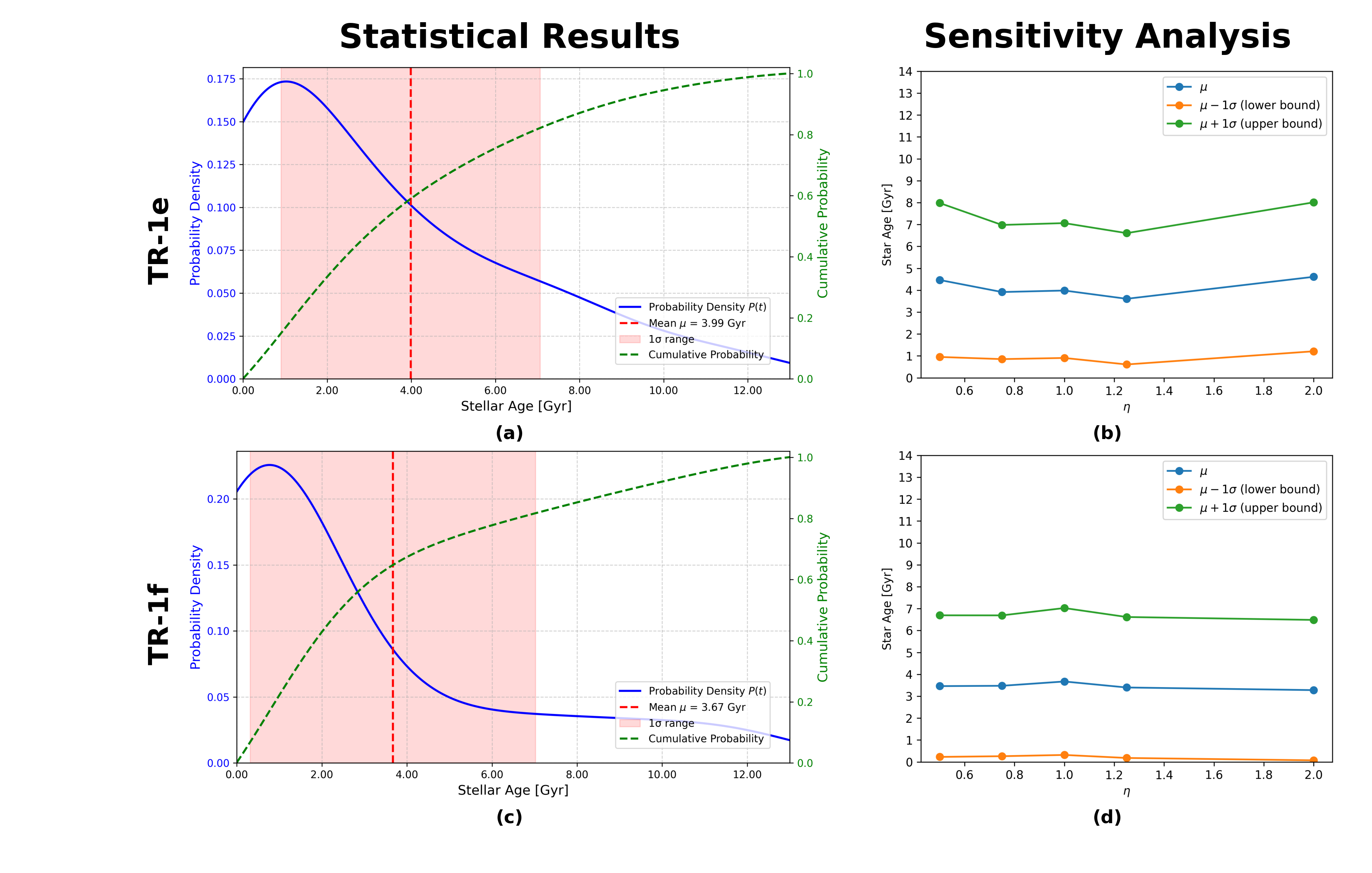}
    \caption
{
TR-1 stellar age inference from TR-1e and TR-1f: results and sensitivity.
(a) \emph{Top-left}: Statistical results for TR-1 stellar ages inferred from TR-1e.
      The blue curve shows the probability density $p(t)$; the green dashed curve is the cumulative probability;
      the vertical red dashed line marks the mean value, $\mu = 3.99\,\mathrm{Gyr}$;
      the pink shaded region denotes the $1\sigma$ credible interval $[0.91,\,7.06]\,\mathrm{Gyr}$.
(b) \emph{Top-right}: Sensitivity of the TR-1e inference to the phase space scale $\eta$;
      the blue, orange, and green curves trace $\mu$, the $1\sigma$ lower bound, and the $1\sigma$ upper bound, respectively.
(c) \emph{Bottom-left}: Statistical results for TR-1 stellar ages inferred from TR-1f (same format as panel~a).
      The mean value is $\mu = 3.67\,\mathrm{Gyr}$ with a $1\sigma$ interval $[0.32,\,7.03]\,\mathrm{Gyr}$.
(d) \emph{Bottom-right}: Sensitivity of the TR-1f inference to $\eta$ (same format as panel~b).
  Overall, the TR-1e and TR-1f resluts are consistent, and the $1\sigma$ bounds show weak dependence on $\eta$ except for modest variations in the upper bound.
}
    \label{fig_TR1eandf}
\end{figure}

% \textcolor{red}{
We estimate the age of TR-1e and TR-1f following exactly the same procedure as for GJ 486.
The resulting mean values are shown in Figures \ref{fig_TR1eandf} (a) and (c).
For consistency with \citet{2017ApJ...845..110B}, we report $1\sigma$ uncertainties based on the 16–84\% credible intervals.
As noted above (Section \ref{age}), the mean value does not coincide with the maximum probability density.
For TR-1e and 1f, the mean ages are 3.99 and  3.67 Gyr with a \(1\sigma\) interval of [0.91, 7.06] and [0.32, 7.03] Gyr, respectively.
In addition, our phase space values are the same as those of GJ 486b (\(\Delta M_p=0.5M_\oplus, \Delta a=0.005\text{ AU}\)), so we also analyzed the impact of the selection of phase space on the results, as shown in Figure \ref{fig_TR1eandf} (b) and (d).
% }

% \textcolor{red}{
Figure \ref{fig_TR1eandf} (b) shows the relationship between the inferred age of TR-1e and the phase space parameter $\eta$.
The mean and lower bound of the \(1\sigma\) credible interval exhibit only weak dependence on variations in \(\eta\), clustering around 4 Gyr and 1 Gyr, respectively.
In contrast, the upper bound of the \(1\sigma\) interval is slightly sensitive to \(\eta\), though it remains concentrated within 6–8 Gyr.
Figure \ref{fig_TR1eandf} (d) shows the corresponding results for TR-1f.
In this case, all summary statistics, including the mean as well as the lower and upper limits \(1\sigma\), are almost insensitive to \(\eta\), indicating a weak dependence of inference on the phase space selection.
% }

% \textcolor{red}{
Previous studies have reported broad age ranges for TRAPPIST-1.
\citet{2015ApJ...810..158F} obtained 0.5–10 Gyr and \citet{2017NatAs...1E.129L} reported 3–8 Gyr, both broadly suggesting an old star.
In contrast, several works based on the strength of its nonthermal magnetic emission have suggested that TRAPPIST-1 could be comparatively young \citep{2017A&A...599L...3B, 2017MNRAS.469L..26O}.
To address this discrepancy, \citet{2017ApJ...845..110B} performed a comprehensive assessment of multiple age diagnostics.
They showed that indicators such as position in the color–absolute magnitude diagram, mean density, and surface gravity, while superficially suggestive of youth, either suffer from substantial uncertainties or conflict with the observed density, and therefore do not robustly support a young age.
In addition, the lithium abundance and rotation period provide lower limits of >0.2 Gyr and >0.3 Gyr, respectively, and the activity relative to typical M8 dwarfs is more consistent with an older star in the solar neighborhood.
Stronger constraints arise from metallicity and kinematics: the metallicity-based inferences yield 2.2–9.2 Gyr (SPOCS, Spectroscopic
Properties of Cool Stars, sample) and 3.2–9.2 Gyr (GCS, Geneva-Copenhagen Survey, sample), while kinematic analyses give 3.9–10.5 Gyr (V velocity with GCS sample), 8.1–11.9 Gyr (dispersion simulation with GCS sample prior), and 6.4–11.6 Gyr (dispersion simulation with heating losses prior).
Synthesizing these diagnostics, they recommended a concordance age range of 5.4–9.8 Gyr.
% }

\begin{figure}
    \centering
    \includegraphics[width=1.0\linewidth]{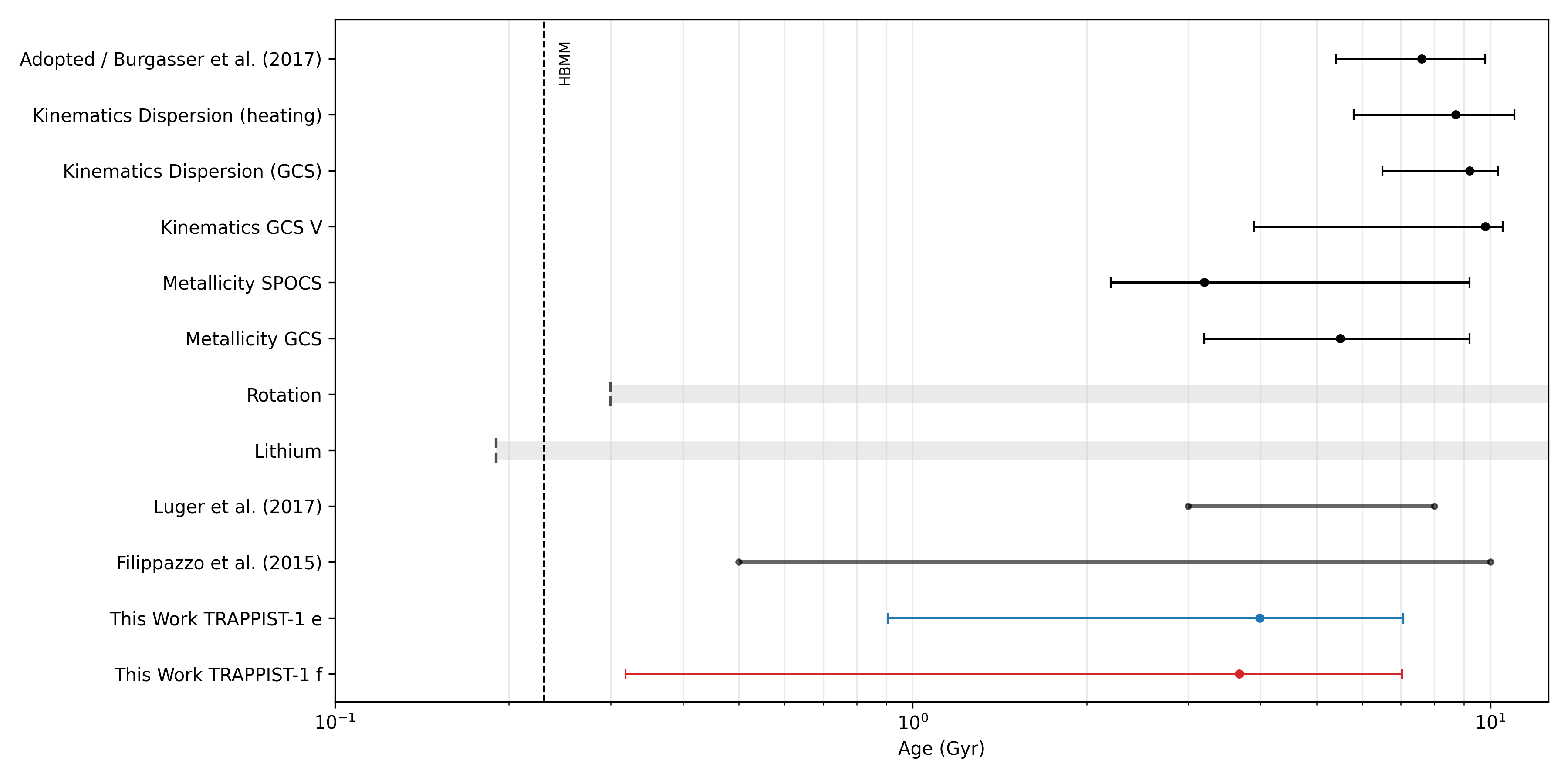}
    \caption
{
Comparison of TR-1 stellar age estimates from the literature and this work (logarithmic age axis).
Horizontal error bars show published age ranges; filled markers denote central values when reported.
From top to bottom: the “Adopted’’ concordance age of \citet{2017A&A...599L...3B};
kinematic constraints from velocity dispersion simulation with priors based on heating losses, from velocity dispersion simulation with priors based on GCS sample, and from the V velocity with GCS sample;
metallicity-based constraints using the SPOCS and GCS samples;
rotation and lithium, plotted as lower-limit indicators (semi-transparent bands starting at $\gtrsim 0.3$~Gyr and $\gtrsim 0.2$~Gyr, respectively);
the ranges from \citet{2017NatAs...1E.129L} and \citet{2015ApJ...810..158F};
and our inferences using TR-1e (blue line) and TR-1f (red line).
Our results are shown as mean values with $1\sigma$ credible intervals:
TR-1e $\mu = 3.99\,\mathrm{Gyr}$, $[0.91,\,7.06]\,\mathrm{Gyr}$;
TR-1f $\mu = 3.67\,\mathrm{Gyr}$, $[0.32,\,7.03]\,\mathrm{Gyr}$.
All literature ranges are shown as given in the original sources; when only limits are available, they are displayed as shaded lower-limit bands.
}
    \label{fig_TR1}
\end{figure}

% \textcolor{red}{
We independently analyze TR-1e and TR-1f and compare our results with those of \citet{2017ApJ...845..110B}, as shown in Figure \ref{fig_TR1}.
Our estimates place the mean ages inferred from both planets at around 4 Gyr, with close agreement between e and f.
The uncertainty interval for e is significantly narrower than for f, primarily because the uncertainty about the water content of e (96.42 TO) is substantially smaller than for f (135.65 TO), which results in a more robust age constraint for e.
Compared to previous work, our mean estimates are consistent with the metallicity-based results of \citet{2017ApJ...845..110B}, though slightly smaller than the kinematic inferences.
In general, our method, in agreement with \citet{2017ApJ...845..110B}, supports the conclusion that TRAPPIST-1 is an old star.
% }

% \subsection{\textcolor{red}{Verifying Dataset Independence of the Age Inference}}
% \label{subsec:VerifyIndenpendence}

\subsection{Verifying Dataset Independence of the Age Inference}
\label{subsec:VerifyIndenpendence}

\begin{figure}
    \centering
    \includegraphics[width=0.5\linewidth]{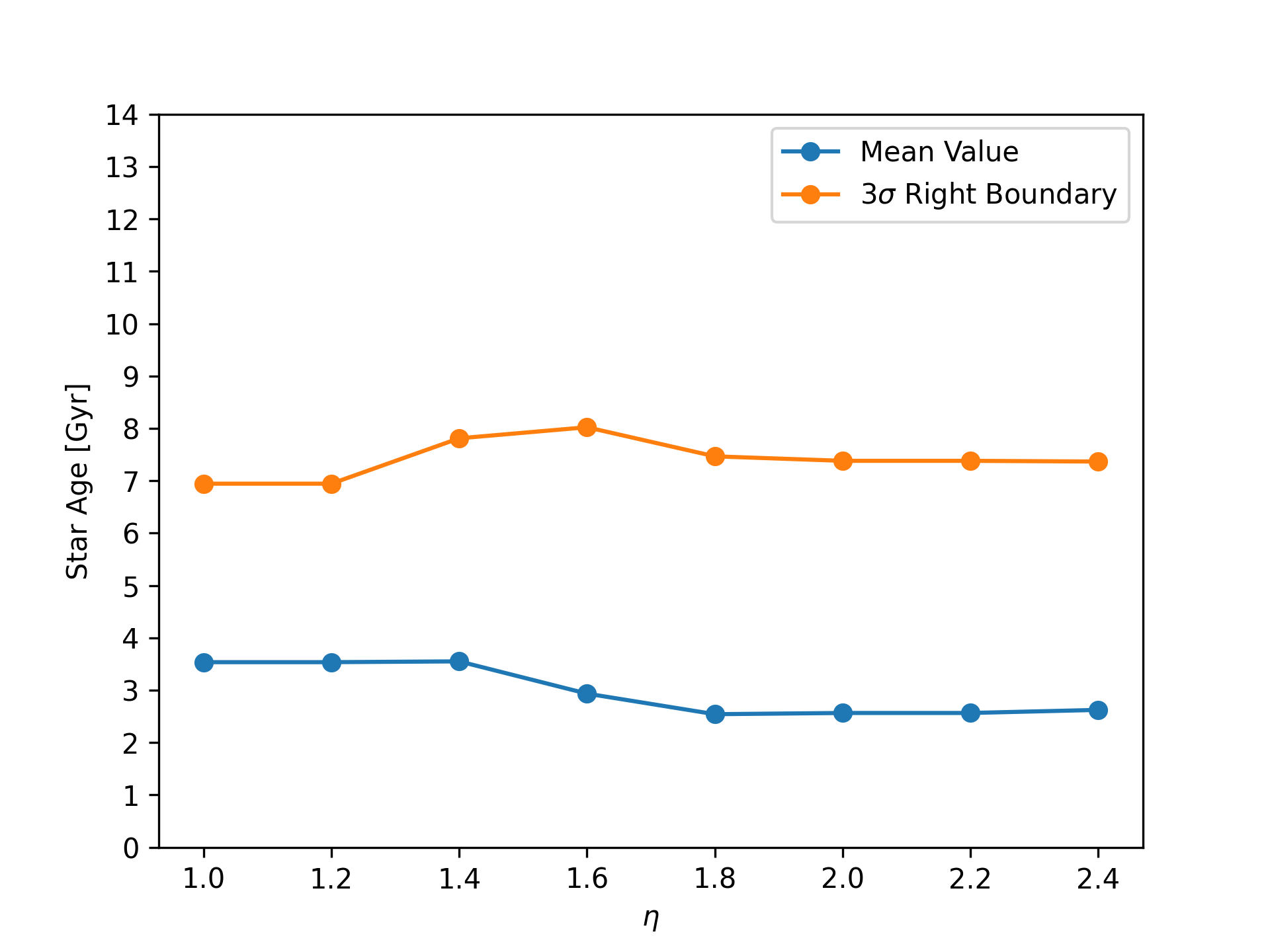}
    \caption
{
Sensitivity of the GJ~486 age inference to the phase space parameter \(\eta\) using the NGPPS dataset.
The blue line indicates the mean value of the stellar age.
The orange first indicates the \(3\sigma\) right boundary of the stellar age.
}
    \label{fig_GJ486newDataset2}
\end{figure}

\begin{figure}
    \centering
    \includegraphics[width=0.8\linewidth]{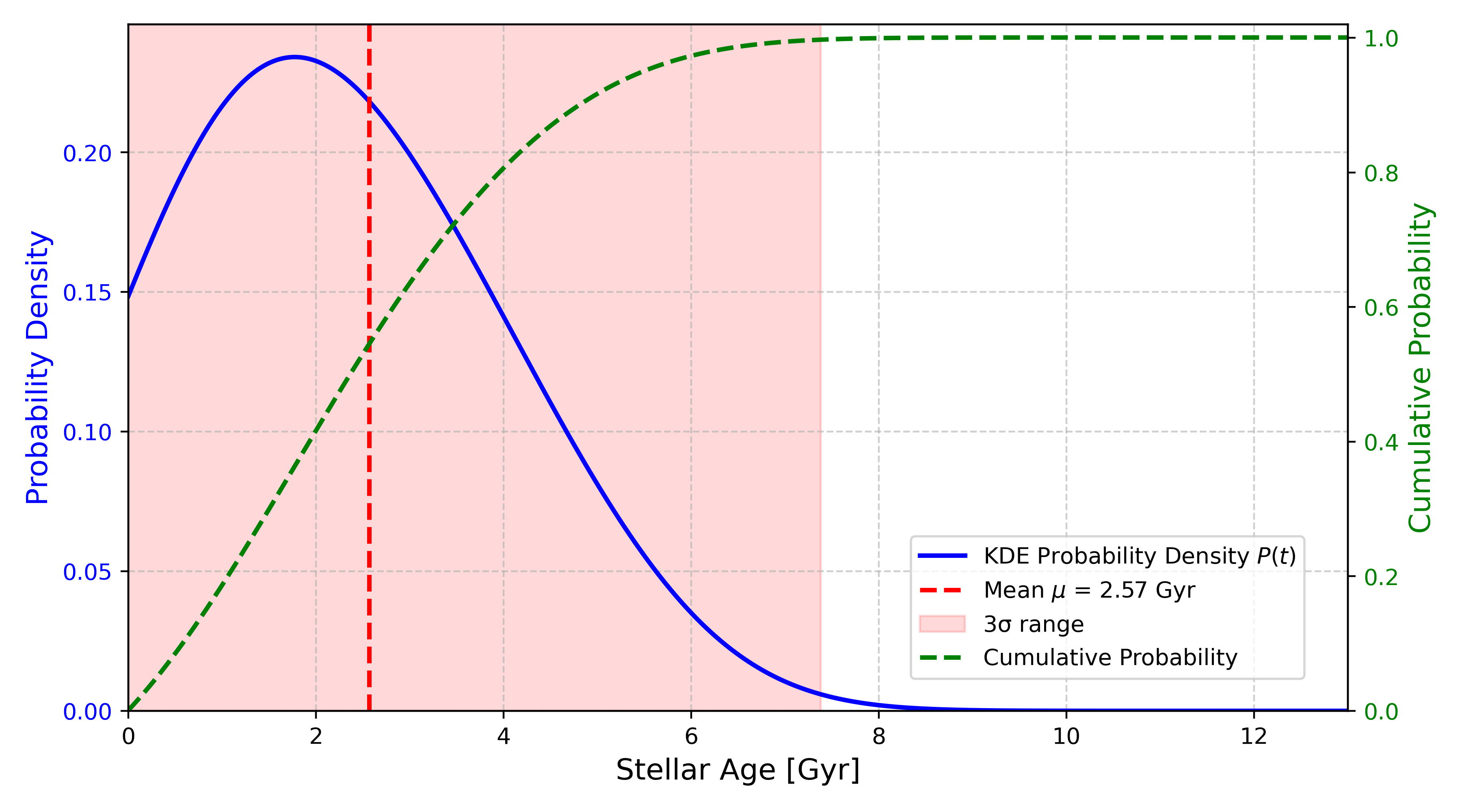}
    \caption
{
Statistical results for stellar ages using the NGPPS dataset.
The red line indicates the expectation of these data, with a value of 2.57 Gyr.
The pink area indicates the \(3\sigma\) range of these data, with a range of \([0, 7.38]\) Gyr.
}
    \label{fig_GJ486newDataset}
\end{figure}

% \textcolor{red}{
To assess whether our inference is sensitive to the planet formation dataset, in this part we additionally used a alternative planet formation dataset, the New Generation Planetary Population Synthesis (NGPPS) \citep{2005A&A...434..343A, 2012A&A...547A.111M, 2014prpl.conf..691B, 2021A&A...656A..69E, 2021A&A...656A..70E}, to estimate the age of GJ 486b.
For stars of mass \(0.3M_\odot\), the NGPPS dataset contains fewer than 50,000 formation cases, whereas the \citet{2022NatAs...6.1296K} dataset includes more than 200,000.
Therefore, we need to be cautious when choosing the size of the phase space.
Here, we still use the same phase space as before as the phase space selection when \(\eta=1.0\).
Figure \ref{fig_GJ486newDataset2} shows the relationship between \(\eta\) and the mean value and the right boundary of \(3\sigma\) using the NGPPS dataset.
We find that the mean value decreases significantly around \(\eta=1.4-1.8\).
This is because the small number of samples results in the phase space containing only a few samples.
Therefore, we choose \(\eta=2.0\) as the phase space for this case.
Figure \ref{fig_GJ486newDataset} shows the statistical results for GJ 486 using the NGPPS dataset.
The mean value is 2.57 Gyr and the \(3\sigma\) right boundary is 7.38 Gyr.
This is almost identical to the results of the \citet{2022NatAs...6.1296K} dataset, with a mean value of 2.90 Gyr.
Combining the results from the NGPPS and \citet{2022NatAs...6.1296K} datasets, we demonstrate that our results from the stellar age analysis of GJ 486 are insensitive to the planet formation dataset.
% }

% \subsection{\textcolor{red}{Effect of Protoplanetary Disk Lifetime}}
\subsection{Effect of Protoplanetary Disk Lifetime}

% \textcolor{red}{
In this study, we adopt a protoplanetary disk lifetime of 5 Myr, motivated by observational constraints indicating characteristic lifetimes of 4.2-5.8 Myr \citep{2014A&A...561A..54R} and consistent with previous simulation setups \citep{2016ApJ...829...63S, 2021AsBio..21.1325B, 2024PSJ.....5..137G}.
Some studies also suggest lifetimes around 3 Myr \citep{2009AIPC.1158....3M, 2018ApJ...859...21A}.
In addition, the protoplanetary disk completely dissipated around 10 Myr \citep{2014A&A...561A..54R}.
Accordingly, in addition to our choice, we also explore disk lifetimes of 3 Myr and 10 Myr to assess the sensitivity of the inferred age of GJ 486b to this assumption of the disk lifetime.
% }

\begin{figure}
    \centering
    \includegraphics[width=1.0\linewidth]{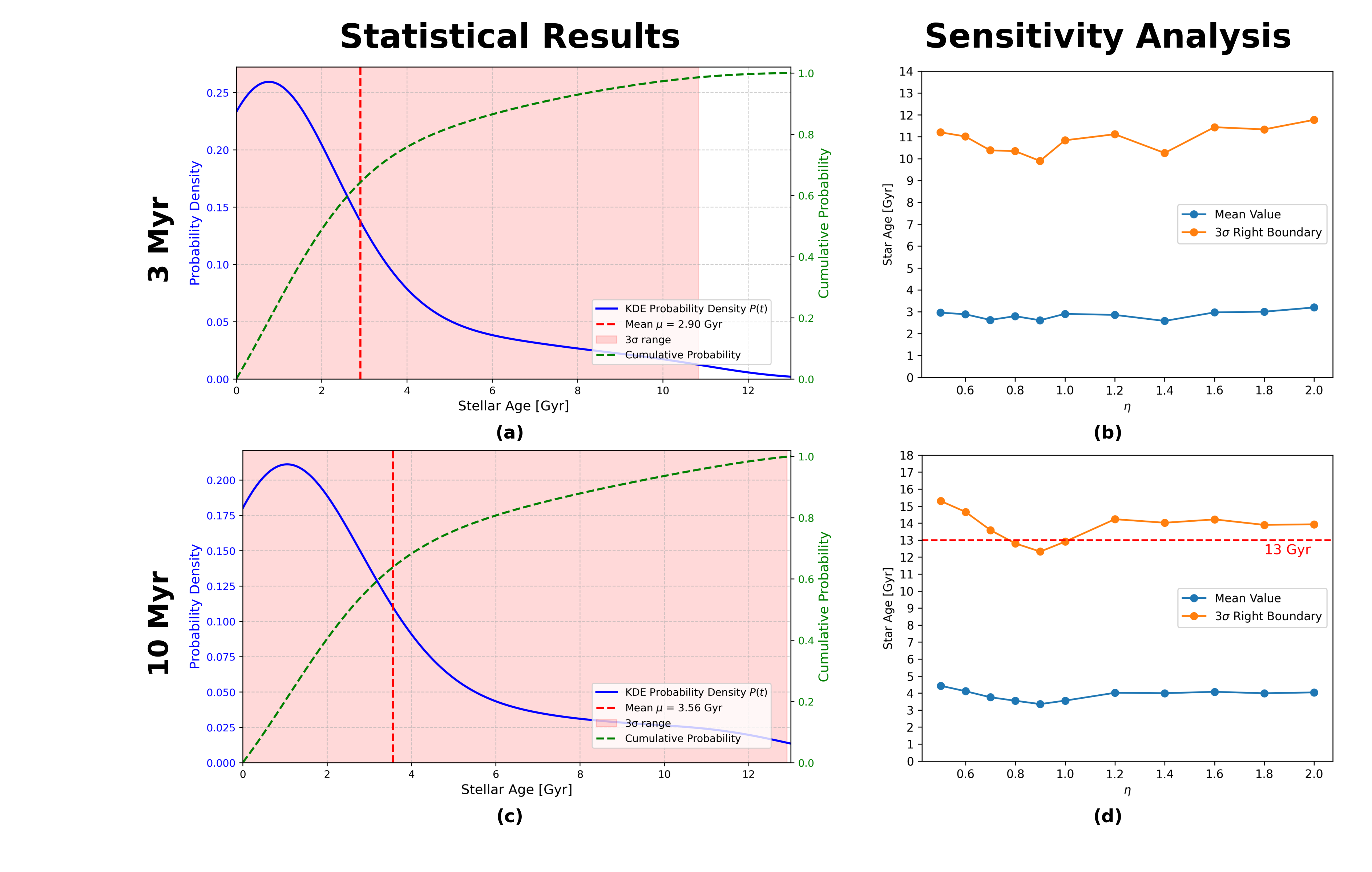}
    \caption
{
GJ 486 stellar age inference from protoplanetary disk lifetimes of 3 and 10 Myr: results and sensitivity.
(a) \emph{Top-left}: Statistical results for GJ~486 stellar ages inferred with a protoplanetary disk lifetime of 3 Myr.
      The blue curve shows the probability density $p(t)$; the green dashed curve is the cumulative probability;
      the vertical red dashed line marks the mean value, $\mu = 2.90\,\mathrm{Gyr}$;
      the pink shaded region denotes the $1\sigma$ credible interval $[0,\,10.84]\,\mathrm{Gyr}$.
(b) \emph{Top-right}: Sensitivity of the 3 Myr inference to the phase space scale $\eta$;
      the blue and orange curves trace the posterior mean and the $3\sigma$ upper boundary, respectively.
(c) \emph{Bottom-left}: Statistical results for GJ~486 stellar ages inferred with a protoplanetary disk lifetime of 10 Myr (same format as panel~a).
      The mean value is $\mu = 3.56\,\mathrm{Gyr}$ with a $1\sigma$ interval $[0,\,12.91]\,\mathrm{Gyr}$.
(d) \emph{Bottom-right}: Sensitivity of the 10 Myr inference to $\eta$ (same format as panel~b).
      The red dashed line marks 13 Gyr, the upper limit of the age of the Universe.
      For the 10 Myr case, the statistical $3\sigma$ range almost coincides with the physical limit of 13 Gyr and is therefore not meaningful.
}
    \label{fig_GJ4863and10Myr}
\end{figure}

% \textcolor{red}{
The resulting mean valus are shown in Figure \ref{fig_GJ4863and10Myr} (a) and (c).
As shown above (Section \ref{age}), the mean value does not coincide with the maximum probability density.
For the case of a protoplanetary disk lifetime of 3 Myr, the mean age is 2.90 Gyr with a \(3\sigma\) interval of \([0,\,10.84]\) Gyr, which is almost consistent with the 5 Myr case.
For the case of a protoplanetary disk lifetime of 10 Myr, the mean age is 3.56 Gyr with a \(3\sigma\) interval of \([0,\,12.91]\) Gyr.
The 3 Myr results in Figure \ref{fig_GJ4863and10Myr} (a) and (b) are nearly indistinguishable from the 5 Myr results (see Figures \ref{fig3} and \ref{fig5}).
This is because the change in the inversion result is so small that none of the planet formation samples is reassigned to different inversion regions; consequently, the differences are driven solely by the area of the inversion regions rather than by any redistribution of samples.
Compared with the 5 Myr result, the 10 Myr case yields a larger mean value, while the \(3\sigma\) range becomes essentially uninformative due to its proximity to the 13 Gyr physical limit.
The mean value is essentially insensitive to the disk lifetime for the more common scenarios with lifetimes shorter than 5 Myr.
For disks with a lifetime dissipation time of \(\sim10\ \text{Myr}\), the age inference increases with disk lifetime, but for GJ 486 star, even in this extreme case, it just varies about 0.6 Gyr.
In addition, our phase space values are the same as those of protoplanetary disk lifetime of 3 Myr (\(\Delta M_p=0.5M_\oplus, \Delta a=0.005\text{ AU}\)), so we also analyzed the impact of the selection of phase space on the results, as shown in Figures \ref{fig_GJ4863and10Myr} (b) and (d).
% }

\subsection{Effect of Magma Ocean Phase}\label{subsec:effect_magma}

In this study, we use the water loss formula provided by \citet{2024PSJ.....5..137G} in VPLanet to calculate water escape during planetary evolution and the detailed settings are in the Section \ref{subsubsec:atmospheric}.
We ignore the effect of the magma ocean phase on planetary water escape \citep{2021AsBio..21.1325B, 2023MNRAS.526.6235M}.
Fortunately, the magma ocean duration of GJ 486b was discussed using internal structural models by \citet{2025ApJ...981...80S}.

% \textcolor{red}{
To quantitatively test this assumption for GJ 486b, we fit a power law to the MO durations reported by \citet{2025ApJ...981...80S} for initial water contents \(W=\{0.2,\,1.0,\,3.0,\,10,\,20\}\,\mathrm{TO}\) and corresponding durations \(t_{\rm MO}=\{0.02,\,0.09,\,0.3,\,0.93,\,1.82\}\,\mathrm{Myr}\).
In log–log space, an ordinary least-squares fit yields
% }
\begin{equation}
% \textcolor{red}{
t_{\rm MO} \;=\; a\, W^{\,b},
% }
\end{equation}
% \[
% \textcolor{red}{
% t_{\rm MO} \;=\; a\, W^{\,b},
% \qquad a = 0.0960~\mathrm{Myr}\; (95\%~\mathrm{CI}:[0.0883,\,0.1044]),\quad
% b = 0.9860\; (95\%~\mathrm{CI}:[0.9418,\,1.0301]),
% }
% \]
% \textcolor{red}{
The best-fit parameters are $a = 0.0960~\mathrm{Myr}\; (95\%~\mathrm{CI}:[0.0883,\,0.1044])$, $b = 0.9860\; (95\%~\mathrm{CI}:[0.9418,\,1.0301])$, with \(R^2=0.9994\) and residual standard deviation \(\sigma_{\log}=0.0510\).
The slope \(b\simeq1\) indicates an approximately linear scaling of MO duration with initial water content over the calibrated range.
Extrapolating to \(W=370\,\mathrm{TO}\) gives a median prediction $t_{\rm MO}(370) \;=\; 32.69~\mathrm{Myr}$ with a 95\% confidence interval for the mean response of \([25.96,\,41.16]~\mathrm{Myr}\) and a 95\% prediction interval of \([24.66,\,43.33]~\mathrm{Myr}\).
Even the upper bound of these intervals remains below our model time step of \(50~\mathrm{Myr}\).
We therefore treat the magma ocean phase as shorter than a single time step and conclude that its effect on the inferred initial water inventory and stellar age constraints is negligible in our framework.
% }

\subsection{Compared with Previous Stellar Age Estimates} \label{subsec:compare}

In Section \ref{age}, we proposed an approach to estimate the stellar age of GJ 486 by combining inversion results with a planetary formation dataset.
Our inference yields a posterior distribution with a stellar age of $2.90^{+2.47}_{-2.27}~\mathrm{Gyr}$ and a 99.7\% credible upper limit of 10.85 Gyr.
We now compare our results with previous estimates reported in the literature.
\citet{2021Sci...371.1038T} and \citet{2022A&A...665A.120C} found a stellar rotation period of 130 days and 49.9 days for GJ 486, which corresponds to age ranges of $3.0$--$12.6$ and $0$--$8.8$\,Gyr based on age–rotation relations, respectively.
% In contrast, \citet{2022A&A...665A.120C} reported a significantly shorter rotation period of $49.9 \pm 5.5$ days, implying a younger stellar age of $0$--$8.8$\,Gyr using the similar method.
In addition, \citet{2024A&A...689A..48D} employed stellar activity indicators derived from the spectrum of GJ 486 (spanning 5 to 1700 \AA) estimated a lower limit of $6.6$\,Gyr for the stellar age.
Our inferred age range for GJ 486 closely matches that reported by \citet{2022A&A...665A.120C}, while also substantially overlapping with the results of \citet{2021Sci...371.1038T} and \citet{2024A&A...689A..48D}, but with a smaller upper bound compared to the latter two studies.
Due to the uncertainties in measuring rotational periods, better observations are required to constrain our proposed methods.

\begin{figure}
    \centering
    \includegraphics[width=0.5\linewidth]{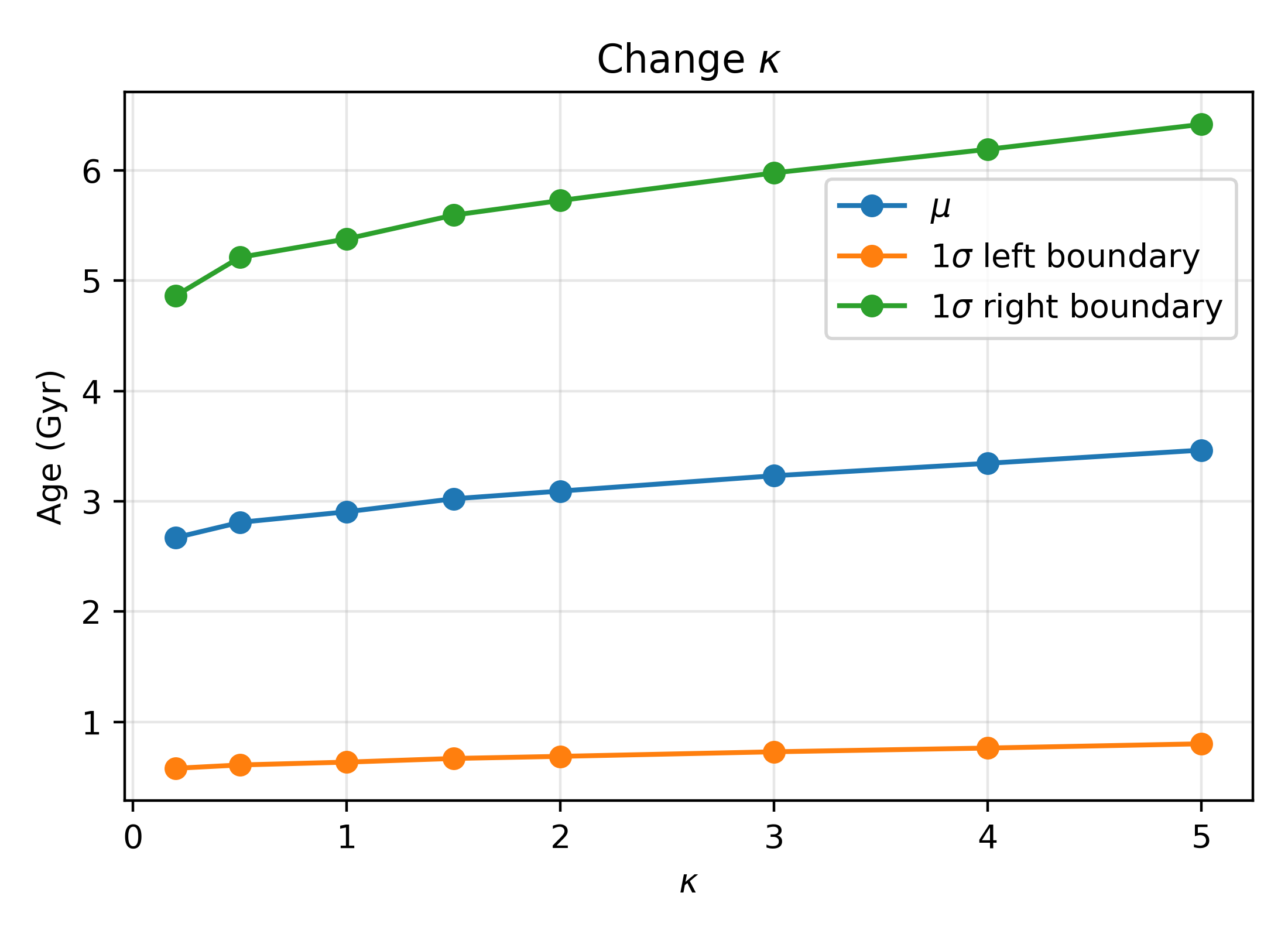}
    \caption
{
The changes of the mean value of the age and uncertainty with the variations of water content uncertainty.
The blue, orange and green lines represent the mean age \(\mu\), the \(1\sigma\) left boundary and the \(1\sigma\) right boundary, respectively.
$\kappa$ represents the upper/lower boundaries of GJ 486b water content multiplied/divided by $\kappa$, respectively.
}
    \label{fig_kappa_and_bias}
\end{figure}

In addition, we discussed the error in the stellar age resulted by the error in the current water content.
In fact, the water content we used in the model differed by 1000 times.
Figure \ref{fig_kappa_and_bias} shows the changes of the mean value of the age and uncertainty with the variations of water content.
As the uncertainty of water content decreases, the accuracy of our method for stellar age is also improved, Vice versa.
This indicates that as the observational accuracy of water improves, the accuracy of our method for stellar age estimation also improves.
In the future, with improvements in both observations and models, this inversion can provide an independent method for estimating stellar ages.

\section{Conclusion} \label{sec:con}

In this study, we modeled the atmospheric evolution of GJ~486b under intense XUV irradiation,  and used the current water content to limit the mass of primordial hydrogen-rich atmosphere and water inventory. By scanning a broad parameter space in initial water content, we obtained families of initial conditions consistent with the same present-day water reservoir. We found that the results of inversion exhibits a clear degeneracy between the primordial hydrogen-rich atmosphere and water retention. Without a hydrogen-rich atmosphere, reproducing the current water content requires a substantially larger primordial water inventory, whereas introducing even a moderate H/He envelope rapidly lowers the required initial water by delaying the onset of efficient water loss.

We also showed that the inferred primordial state is highly age-dependent. To move beyond purely ``allowed regions'' in parameter space, we incorporated a planet formation prior to determine the age of GJ~486b. This provides an additional physical filter on plausible primordial (water, H/He) combinations and, when combined with the inversion results, yields an expected stellar age of $2.90^{+2.47}_{-2.27}$~Gyr with a 99.7\% interval of 0--10.85~Gyr. We further tested the robustness of the age-inference framework against several modeling choices, including the use of an alternative formation dataset, variations in disk lifetime, and reasonable changes in the selection of the phase-space. These tests support the stability of the inferred mean age at the $\sim$Gyr level.
Benchmarking on TRAPPIST-1 e and f gives mutually consistent ages and broadly agrees with literature constraints, indicating that the method can be further
inspected in other potentially possible planets.

\begin{acknowledgments}

We thank the reviewer for their constructive comments and
suggestions, which helped to improve the quality of this work.
We would also like to thank Haihao Shi for helpful discussions on topics related to this work.

We gratefully acknowledge Tadahiro Kimura and Masahiro Ikoma for kindly providing the dataset for this study.
We also gratefully acknowledge the Data \& Analysis Center for Exoplanets for sharing the NGPPS dataset with us.
This publication makes use of The Data \& Analysis Center for Exoplanets (DACE), which is a facility based at the University of Geneva (CH) dedicated to extrasolar planets data visualisation, exchange and analysis. DACE is a platform of the Swiss National Centre of Competence in Research (NCCR) PlanetS, federating the Swiss expertise in Exoplanet research. The DACE platform is available at https://dace.unige.ch.

This work is supported by the National Natural Science
Foundation of China (Grant No. 12433009, 12288102 and 11973082). We thank the support by YUNNAN ADMINISTRATION OF FOREIGN EXPERTS
AFFAIRS, grant No.202505AO120026. We also acknowledge support from the National Key R\&D Program of China (Grant No.
2021YFA1600400/2021YFA1600402) and International Centre of Supernovae (ICESUN), Yunnan Key Laboratory of Supernova Research (No. 202505AV340004).

\end{acknowledgments}


\vspace{5mm}
% \facilities{HST(STIS), Swift(XRT and UVOT), AAVSO, CTIO:1.3m,


\software{astropy \citep{2013A&A...558A..33A,2018AJ....156..123A},
          VPLanet \citep{2020PASP..132b4502B},
          NumPy \citep{harris2020array},
          SciPy \citep{2020SciPy-NMeth},
          Matplotlib \citep{Hunter:2007},
          scikit-learn \citep{scikit-learn},
          dace-query (To download the NGPPS dataset) \citep{2005A&A...434..343A, 2012A&A...547A.111M, 2014prpl.conf..691B, 2021A&A...656A..69E, 2021A&A...656A..70E}
          }



% APPENDIX BEGIN

\appendix

\section{Stellar Age Estimation by Planetary Formation Dataset} \label{app1}

In this appendix, we design a method to estimate stellar ages by combining water content inversion results with planetary formation datasets, and obtain an age estimate for the star GJ 486.

Let \( A \) denote the set of all planetary formation data points in the \( (M_w, M_\text{H/He}) \) plane, where each point corresponds to a simulation with a specific initial water content and atmospheric mass.
In Section \ref{subsec:DatasetAnalysis}, we described in detail how to obtain this set \( A \).
For simplicity, since there are no planets in the dataset with exactly the same semi-major axis and mass as GJ 486b, we define a phase space \((a_0\pm\Delta a, M_p\pm\Delta M_p)\).
We consider all samples within this phase space \((a_0\pm\Delta a, M_p\pm\Delta M_p)\) to be highly similar to GJ 486b.
Here, we set the phase space parameters \(\Delta a=0.005\)AU and \(\Delta M_p=0.5M_\oplus\).
In the dataset, planets with semi-major axes and planetary masses located within this phase space, whose initial water content and initial atmospheric mass constitute set \(A\).
For each stellar age \( t \), we define a corresponding region \( S(t)\), derived from the inversion results.

For example, in Figure \ref{fig3_res} we show the regions \(S(3.7\text{ Gyr})\), \(S(7.8\text{ Gyr})\) and \(S(10.5\text{ Gyr})\) for stellar ages \(t\) of 3.7, 7.8 and 10.5 Gyr.
We then define the probability \( P(t) \) to be proportional to the ratio between the number of planetary formation points located within \( S(t) \), and the area of that region:

\begin{equation}
    P(t) \propto \frac{|A \cap S(t)|}{\mathrm{Area}(S(t))} \label{eqp}
\end{equation}

Where \( |A \cap S(t)| \) denotes the number of data points in \( A \) that fall inside the region \( S(t) \), and \( \mathrm{Area}(S(t)) \) is the area of that region.

To transform the discrete probabilities \(P(t_i)\) obtained from Equation~\ref{eqp} into a continuous, smooth probability density function (pdf), we adopt a weighted Gaussian kernel density estimator.  Let \(\{t_i\}_{i=1}^{N}\) be the set of stellar age grid points at which Equation~\ref{eqp} is evaluated, and define the normalized weights:

\begin{equation}
    w_i \;=\; \frac{P(t_i)}{\sum_{j=1}^{N}P(t_j)},
    \qquad
    \sum_{i=1}^{N}w_i = 1
\end{equation}

We employ the standard Gaussian kernel and define the standardized offset:

\begin{equation}
  K(u) = \frac{1}{\sqrt{2\pi}}\exp\!\Bigl(-\tfrac12u^2\Bigr),
  \qquad
  u_i = \frac{t - t_i}{h}
  \label{eq:kernel_offset}
\end{equation}

Where \(u_i\) measures the deviation of an arbitrary evaluation point \(t\) from the \(i\)-th grid node \(t_i\) in units of the bandwidth \(h\).  The bandwidth is then chosen according to Scott’s rule \citep{2015mdet.book.....S}:

\begin{equation}
  h = \sigma_{w}\,N^{-1/5},
  \quad
  \sigma_{w}^{2} = \sum_{i=1}^{N}w_i\,(t_i - \bar t_{w})^{2},
  \quad
  \bar t_{w} = \sum_{i=1}^{N}w_i\,t_i
  \label{eq:bandwidth}
\end{equation}

Subsequently, we introduce the bandwidth scaled kernel:

\begin{equation}
  K_h(x) \;=\;\frac{1}{h}\,K\!\Bigl(\frac{x}{h}\Bigr)
           = \frac{1}{h\sqrt{2\pi}}
             \exp\!\Bigl(-\tfrac12\bigl(\tfrac{x}{h}\bigr)^2\Bigr)
  \label{eq:scaled_kernel}
\end{equation}

So that, for an arbitrary evaluation point \(t\), the contribution of the \(i\)-th grid node \(t_i\) is:

\begin{equation}
  w_i\,K_h(t - t_i)
  = w_i\,\frac{1}{h\sqrt{2\pi}}
    \exp\!\bigl(-\tfrac12\,u_i^2\bigr)
\end{equation}

Summing over \(i=1,\dots,N\) immediately recovers the estimator:

\begin{equation}
    \hat{f}(t) \;=\;
    \sum_{i=1}^{N}\frac{w_i}{h\sqrt{2\pi}}
        \exp\!\Bigl[-\tfrac12\bigl(\tfrac{t-t_i}{h}\bigr)^{2}\Bigr]
    \label{eq:kde}
\end{equation}

Finally, the estimate is renormalized via the trapezoidal‐rule integral to enforce
\(\int \hat f(t)\,\mathrm{d}t = 1\), yielding the fully normalized KDE‐based probability density function.

Beyond furnishing a visually smooth curve, this method offers several tangible advantages.
First, weighting by \(P(t_i)\) retains the full statistical power of the raw inversion counts while automatically correcting for heterogeneous sampling in the \((x,y)\) plane.
This approach ensures that no information is lost through arbitrary thresholding or binning.
Furthermore, Scott’s bandwidth minimises the leading‐order mean integrated squared error, striking a data‐driven balance between bias and variance \citep{silverman1986density}.
This guard against both spurious small‐scale structure in sparsely populated tails and excessive oversmoothing in well‐sampled cores, thereby enhancing the estimator’s accuracy.
The combination of probabilistic consistency, smoothness, and asymptotic optimality makes Equation \ref{eq:kde} furnishes a smooth, differentiable Gaussian approximation to the discrete probabilities \(P(t_i)\), so we can use to constraints on the age of GJ 486.

\begin{figure}
    \centering
    \includegraphics[width=0.8\linewidth]{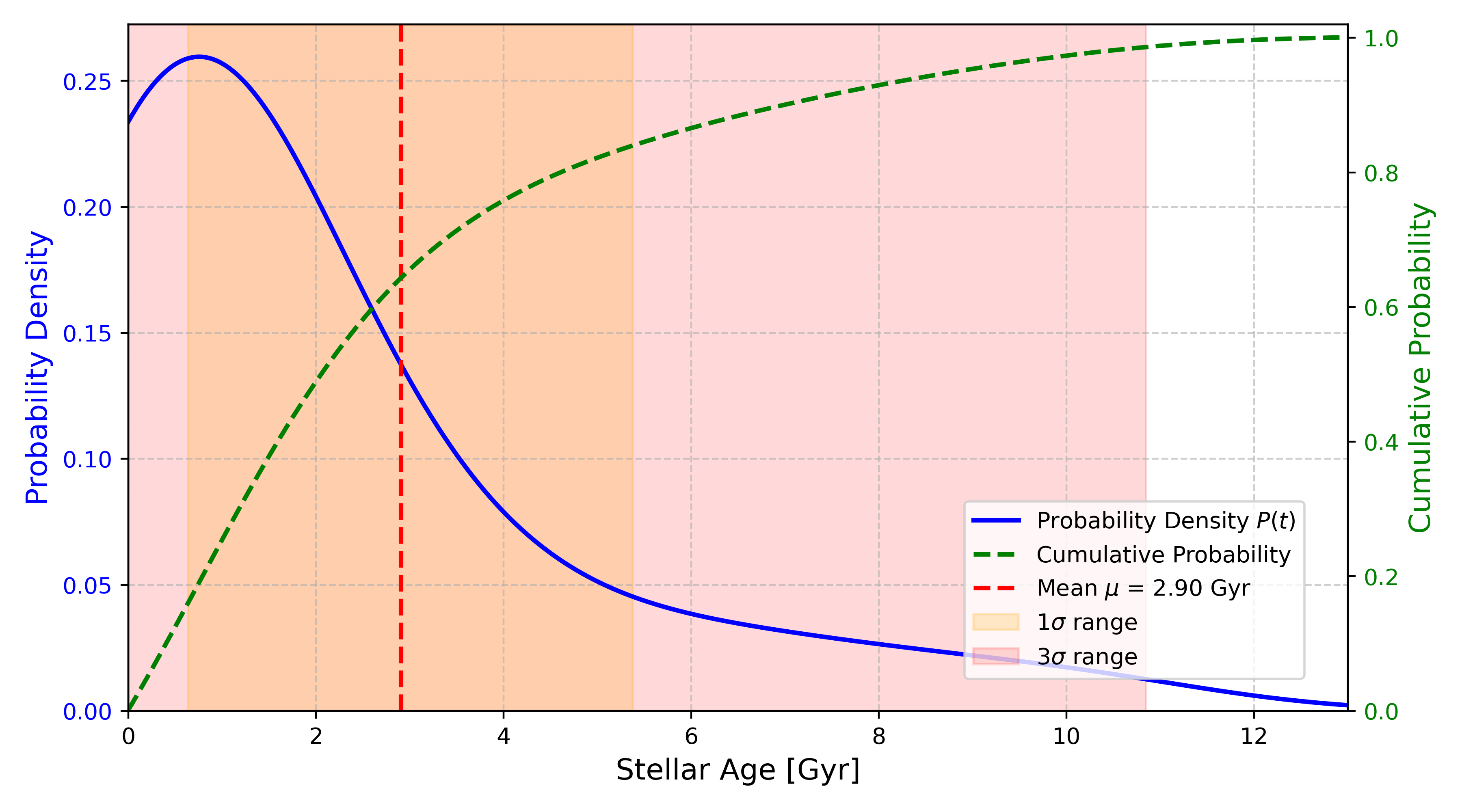}
    \caption
{
Statistical results for stellar ages.
The x-axis indicates the stellar age.
The blue line indicates the probability density, using the left Y-axis.
The green line indicates the cumulative probability, using the right Y-axis.
The red line indicates the expectation of these data, with a value of $2.90^{+2.47}_{-2.27}~\mathrm{Gyr}$.
The pink area indicates the \(3\sigma\) range of these data, with a range of \([0, 10.85]\) Gyr.
}
    \label{fig3}
\end{figure}

Figure \ref{fig3} illustrates the statistical results for stellar ages with the probability defined in Equation \ref{eq:kde}.
The blue curve in Figure \ref{fig3} represents the probability density function $P(t)$ of stellar age $t$, which attains its maximum near $1\, \mathrm{Gyr}$ and subsequently decays rapidly with increasing age.
The green dashed line denotes the cumulative distribution function (CDF), defined as

\begin{equation}
\text{CDF}(t)=\int_0^t P(t')\,\mathrm{d}t',
\end{equation}

which approaches 50\% around 3 Gyr and asymptotically approaching 99.7\% at 10.85 Gyr.
This is consistent with the upper boundary of a 3$\sigma$ interval in a Gaussian-like distribution.
The region where the \(\text{CDF}(t)\) remains below 99.7\% is shaded in pink.

Although the fitting function (as given in Equation \ref{eq:kde}) is formally defined over the entire real axis $t \in (-\infty, +\infty)$, stellar age is physically constrained to be non-negative.
Consequently, the plotted probability density corresponds to the renormalized truncated distribution restricted to the domain of $t > 0$.
This truncation fundamentally alters the statistical properties of the distribution.
In particular, while the location of the maximum of $P(t)$ lies near $1\, \mathrm{Gyr}$, the mean stellar age must be computed as the expectation over the truncated distribution:

\begin{equation}
    \mu = \int_0^{+\infty} t\,P(t)\,\mathrm{d}t.
\end{equation}

The resulting value, $\mu = 2.90\, \mathrm{Gyr}$, is indicated by the red dashed line in Figure \ref{fig3}.
Thus, we conclude that GJ 486 is likely to be a star with an expected age of $2.90^{+2.47}_{-2.27}~\mathrm{Gyr}$ and a \(3\sigma\) range of 0 to 10.85 Gyr based on our inversion and the planetary formation dataset by \citet{2022NatAs...6.1296K} with the phase space of \(\Delta a=0.005\)AU, \(\Delta M_p=0.5M_{\oplus}\).

% In the above analysis, we selected the planet formation dataset with the phase space of \(\Delta a=0.005\)AU, \(\Delta M_p=0.5M_{\oplus}\).
Here we consider the effect of variations in the phase space selection on the stellar age estimates.
For convenience, we use a variable \(\eta\) to compute the stellar ages we obtain using the same method as before when the phase space is selected as \(\Delta a=0.005\eta\)AU, \(\Delta M_p=0.5\eta M_{\oplus}\).
We choose the range 0.5-2.0 for \(\eta\) and discuss the effect of the choice of phase space on the stellar age estimates.

\begin{figure}
    \centering
    \includegraphics[width=0.5\linewidth]{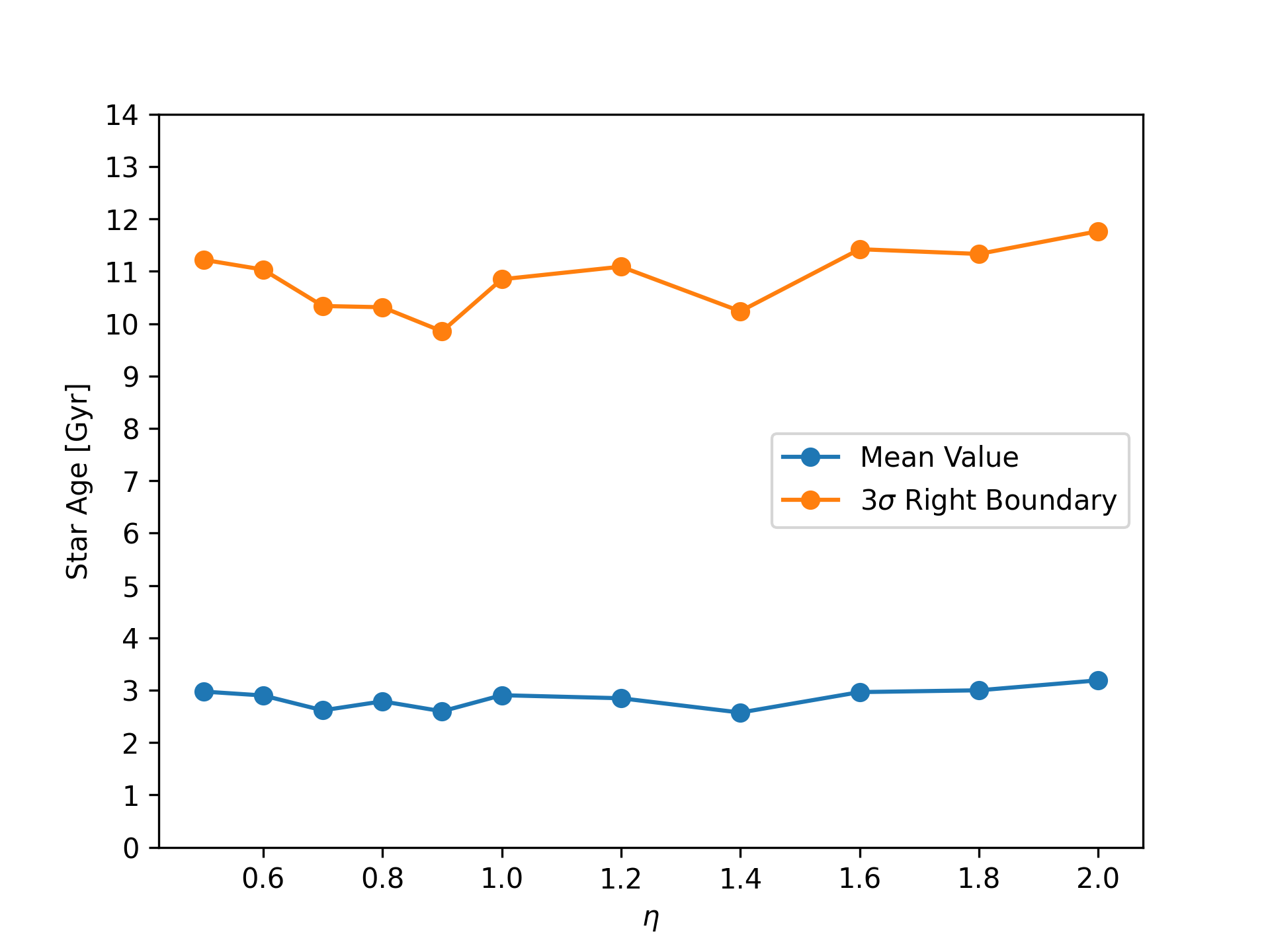}
    \caption
{
Statistical results of stellar ages for different values of \(\eta\).
The blue line indicates the mean value of the stellar age.
The orange first indicates the \(3\sigma\) right boundary of the stellar age.
}
    \label{fig5}
\end{figure}

\bibliography{sample631}{}
\bibliographystyle{aasjournal}



\end{document}